\documentclass[aps,prd,onecolumn,showpacs,superscriptaddress,preprintnumbers,floatfix,reprint]{revtex4-1}
\usepackage{graphicx}
\usepackage{amsmath}
\usepackage{bm}
\usepackage{yhmath}
\usepackage[colorlinks,citecolor=blue,urlcolor=blue,linkcolor=blue]{hyperref}
\usepackage{subfigure}
\usepackage{color}
\usepackage{cases}
\usepackage{subfigure}
\usepackage{dcolumn,booktabs,bm}
\usepackage{slashed}
\usepackage{amsfonts,amssymb,stmaryrd,latexsym,amsmath}
\usepackage{textcomp}
\usepackage{multirow}
\usepackage{cancel}
\usepackage{array}

\allowdisplaybreaks

\begin{document}

\title{Light pseudoscalar meson and doubly charmed baryon scattering lengths with heavy diquark-antiquark symmetry}

\author{Lu Meng}\email{lmeng@pku.edu.cn}
\affiliation{School of Physics and State Key Laboratory of Nuclear
Physics and Technology, Peking University, Beijing 100871, China}

\author{Shi-Lin Zhu}\email{zhusl@pku.edu.cn}
\affiliation{School of Physics and State Key Laboratory of Nuclear
Physics and Technology, Peking University, Beijing 100871,
China}\affiliation{Collaborative Innovation Center of Quantum
Matter, Beijing 100871, China}

\begin{abstract}
We adopt the heavy baryon chiral perturbation theory (HBChPT) to calculate the scattering lengths of  $\phi B_{cc}^{(*)}$ up to $\mathcal{O}(p^3)$, where $\phi$ is the pseudoscalar mesons. The recoil effect and the mass splitting between the spin-$1\over 2$ and spin-$3\over 2$ doubly charmed baryons are included. In order to give the numerical results, we construct the chiral Lagrangians with heavy diquark-antiquark (HDA) symmetry in a formally covariant approach. Then, we relate the low energy constants (LECs) of the doubly charmed baryons to those of $D^{(*)}$ mesons. The LECs for the $\phi D^{(*)}$ scattering are estimated in two scenarios, fitting lattice QCD results and using the resonance saturation model. The chiral convergence of the first scenario is not good enough due to the the large strange quark mass and the presence of the possible bound states, virtual states and resonance. The final results for two scenarios are consistent with each other. The interaction for the $[\pi\Xi^{(*)}_{cc}]^{(1/2)}$, $[K\Xi^{(*)}_{cc}]^{(0)}$, $[K\Omega^{(*)}_{cc}]^{(1/2)}$, $[\eta\Xi^{(*)}_{cc}]^{(1/2)}$, $[\eta\Omega^{(*)}_{cc}]^{(0)}$ and $[\bar{K}\Xi^{(*)}_{cc}]^{(0)}$ channels are attractive. The most attractive channel $[\bar{K}\Xi^{(*)}_{cc}]^{(0)}$ may help to form the partner states of the $D_{s0}^*(2317)$ ($D_{s1}(2460)$) in the doubly heavy sector.
\end{abstract}

\maketitle

\thispagestyle{empty}

\section{Introduction}\label{sec:intro}
The doubly charmed baryon $\Xi_{cc}^{++}$ has been discovered by LHCb Collaboration recently~\cite{Aaij:2017ueg}. The properties and interaction of doubly charmed baryons attract much attention. The mass spectrum of doubly charmed baryons has been investigated in Refs.~\cite{Hyodo:2017hue,Lu:2017meb,Yan:2018zdt,Weng:2018mmf}. The weak and strong decays of the doubly charmed baryons have been extensively studied~\cite{Wang:2017azm,Xiao:2017udy,Wang:2017mqp,Cheng:2018mwu}. The electromagnetic properties and radiative decays of the doubly heavy baryons have been discussed in Refs.~\cite{Li:2017cfz,Meng:2017dni,Li:2017pxa,Bahtiyar:2018vub,Blin:2018pmj}. The possible bound states composed of the doubly charmed baryons were investigated in Refs.~\cite{Meng:2017fwb,Meng:2017udf,Chen:2018pzd,Guo:2017vcf}.

In the large mass limit, the two heavy quarks in the doubly baryons tend to form a compact heavy diquark. The heavy diquark decouples with the light degrees of freedom, which gives rise to the heavy diquark symmetry. The diquark in the doubly heavy baryon is in color $\bar{3}$ representation. Its color dynamics is the same as the single heavy antiquark, if we treat the diquark as a compact object. In this way, the doubly heavy diquark can be related to a singly heavy antiquark~\cite{Savage:1990di}. We will refer this symmetry as the heavy diquark-antiquark (HDA) symmetry. The domain of the validity of the symmetry were discussed a decade ago in Ref.~\cite{Cohen:2006jg}. With this symmetry, the doubly charmed baryons were related to the charmed mesons. The $\Xi_{cc}^{(*)}$ and the $\Omega_{cc}^{(*)}$ are analogous to the $\bar{D}^{(*)}$ and $\bar{D}_s^{(*)}$, respectively. Hu {\it et al.} constructed the chiral Lagrangian with the HDA symmetry and predicted the radiative decays of the doubly charmed baryons with the properties of the $D$ mesons as input~\cite{Hu:2005gf}. 

$D_{s0}^*(2317)$ and $D_{s1}(2460)$ are charmed-strange mesons in the $(0^+, 1^+)$ doublet~\cite{Aubert:2003fg,Besson:2003cp}. Their masses are much lower than the quark model predictions (for a recent review, see Ref.~\cite{Chen:2016spr}). The $S$-wave $DK$ and $D^*K$ channels may play an important role on lowering their masses~\cite{vanBeveren:2003kd,Dai:2003yg,Guo:2006fu,Lang:2014yfa}. The scattering lengths of the $D_{(s)}\phi$ and $D_{(s)}^*\phi$ have been obtained in the chiral perturbation theory (ChPT)\cite{Liu:2009uz,Liu:2011mi,Guo:2009ct,Geng:2010vw,Altenbuchinger:2013vwa} or lattice QCD simulations~\cite{Liu:2012zya,Mohler:2013rwa,Lang:2014yfa,Moir:2016srx,Guo:2018tjx}. 

Considering the HDA symmetry, it is natural to extend the calculation of scattering lengths to the $B^{(*)}_{cc}\phi$ channels, where $B_{cc}$ and $B^{*}_{cc}$ are the spin-$1\over 2$ and spin-$3\over 2$ doubly charmed baryons, respectively. The study of the scattering lengths may give some clues to the possible bound states of the $B^{(*)}_{cc}\phi$ systems. In Ref.~\cite{Guo:2017vcf}, the leading-order scattering lengths of  the $B_{cc}\phi$ systems are calculated in the chiral effective field theory. In Ref.~\cite{Yan:2018zdt}, a set of negative-parity spin-1/2 doubly charmed baryons are predicted from a unitarized version of ChPT combining with HDA symmetry. In this work, we calculate the scattering lengths of the $B_{cc}\phi$ and $B_{cc}^{*}\phi$ in the heavy baryon chiral perturbation theory (HBChPT) to the next-to-next-to-leading order. We extend the formalism of the heavy quark symmetry~\cite{Falk:1991nq} to the HDA symmetry. The Lagrangians with the HDA symmetry are constructed in a relatively covariant form. The numerical results of scattering lengths are given with the HDA symmetry.

The paper is organized as follows. In Sec.~\ref{sec:Tmatrix}, we list the HBChPT Lagrangians involved. With the Lagrangians, we present the analytical expressions of the $T$-matrices at thresholds. In Sec.~\ref{sec:symmetry}, we reconstruct the Lagrangians with the HDA symmetry. The low energy constants (LECs) for the doubly charmed baryons are related to those for the charmed mesons. In Sec.~\ref{sec:num}, we present the numerical results in two scenarios. We give a brief discussion and conclusion in Sec.~\ref{sec:concl}. We give the details about the superfields and integrals in Appendixes~\ref{app:super} and \ref{app:integrals}.
\section{$T$-Matrices at thresholds}\label{sec:Tmatrix}
\begin{figure}[!htp]
	\centering
	\includegraphics[scale=0.4]{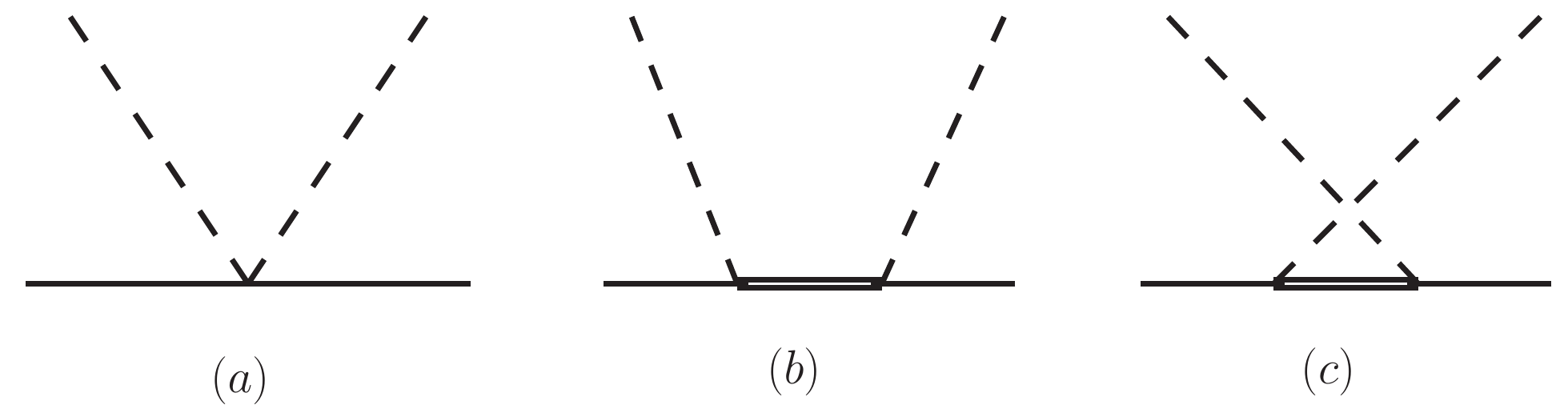}
	\caption{The tree diagrams contributing to the scattering lengths. The single solid lines represent either the spin-$1\over 2$ or the spin-$3\over 2$ doubly charmed baryons. The double solid lines represent both the spin-$1\over 2$ and the spin-$3\over 2$ doubly charmed baryons. }\label{tree}
\end{figure}

\begin{figure*}[!htp]
\centering
\includegraphics[scale=1.0]{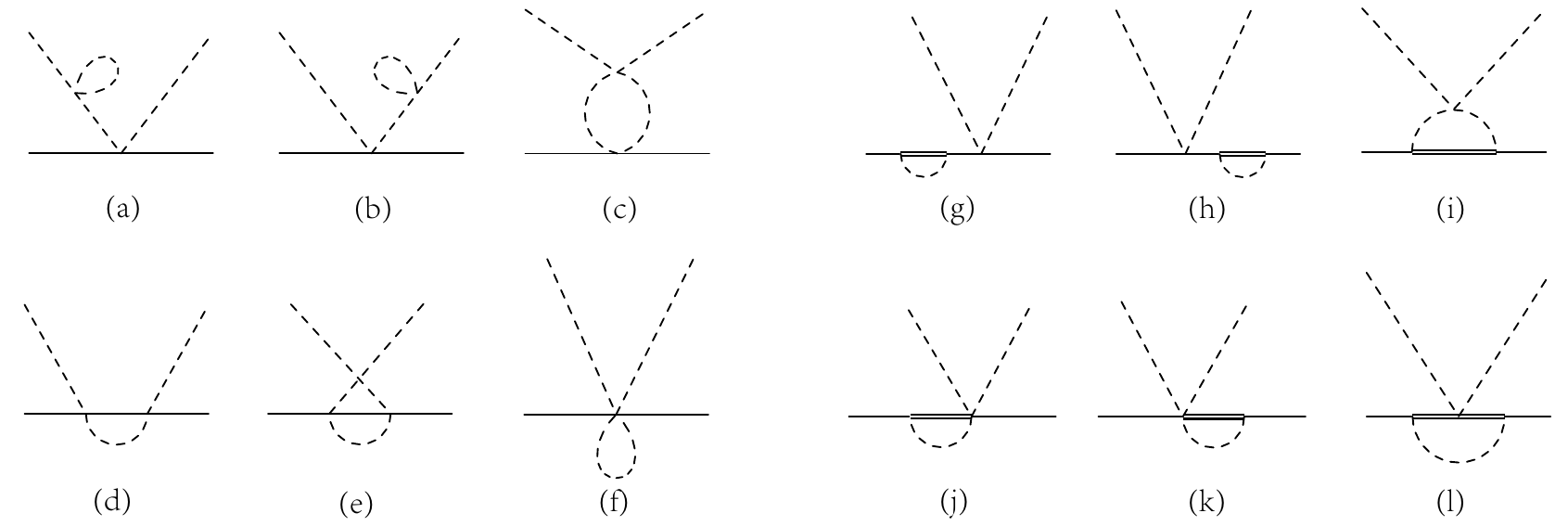}
\caption{The loop diagrams contributing to the scattering lengths at $\mathcal{O}(p^3)$. The loops (a)-(f) in the left panel do not contain the axial vertices. The loops (g)-(l) contain two axial vertices. The dashed lines represent the pseudoscalar mesons. The single solid lines represent either the spin-$1\over 2$ or the spin-$3\over 2$ doubly charmed baryons. The double solid lines represent both the spin-$1\over 2$ and the spin-$3\over 2$ doubly charmed baryons. }\label{loop}
\end{figure*}

The scattering length $a_{\phi B}$ can be derived from the $T$-matrix at threshold,
\begin{equation}
T_{th}=4\pi\left(1+{m_{\phi}\over m_{B}}\right)a_{\phi B}.
\end{equation}
The tree diagrams may contribute to the scattering lengths are shown in Fig.~\ref{tree}.
The loop diagrams contributing to the scattering lengths at $\mathcal{O}(p^3)$ are shown in Fig.~\ref{loop}. Since the spin-$1\over 2$ and spin-$3\over 2$ doubly charmed baryons are degenerate in the heavy diquark limit, we include both of them as the intermediate states. 

In order to calculate the scattering lengths, we construct the chiral Lagrangians order by order. The pseudoscalar mesons can be expressed as,
\begin{eqnarray}
\phi(x)=\left(\begin{array}{ccc}
\pi^{0}+\frac{1}{\sqrt{3}}\eta & \sqrt{2}\pi^{+} & \sqrt{2}K^{+}\\
\sqrt{2}\pi^{-} & -\pi^{0}+\frac{1}{\sqrt{3}}\eta & \sqrt{2}K^{0}\\
\sqrt{2}K^{-} & \sqrt{2}\bar{K^{0}} & -\frac{2}{\sqrt{3}}\eta
\end{array}\right).
\end{eqnarray}
The leading order chiral Lagrangian of pseudoscalar mesons reads,
\begin{equation}
{\cal L}_{\phi\phi}^{(2)}=\frac{F_{0}^{2}}{4}\text{Tr}[\partial_{\mu}U(\partial^{\mu}U)^{\dagger}]+\frac{F_{0}^{2}}{4}\text{Tr}[\chi U^{\dagger}+U\chi^{\dagger}],
\end{equation}
with
\begin{eqnarray}
&U=u^2=\exp(i\phi/F_0),\nonumber \\
&\chi=\text{diag}(m_{\pi}^{2},m_{\pi}^{2},2m_{K}^{2}-m_{\pi}^{2}),
\end{eqnarray}
where $F_0$ is the pion decay constant in the chiral limit.

The ground spin-$1\over 2$ and spin-$3\over 2$ doubly charmed baryons form two triplets in the SU(3) symmetry,
\begin{equation}
B=\left(\begin{array}{c}
\Xi_{cc}^{++}\\
\Xi_{cc}^{+}\\
\Omega_{cc}^{+}
\end{array}\right),\quad B^{*\mu}=\left(\begin{array}{c}
\Xi_{cc}^{*++}\\
\Xi_{cc}^{*+}\\
\Omega_{cc}^{*+}
\end{array}\right)^{\mu},
\end{equation}
where $B$ and $B^*$ are the spin-$1\over 2$ and spin-$3\over 2$ triplets, respectively. In order to construct the Lagrangians of heavy baryons, we define some ``building blocks'',
\begin{eqnarray}
u_{\mu}&=&\frac{i}{2}[u^{\dagger}\partial_{\mu}u-u\partial_{\mu}u^{\dagger}],\\
\Gamma_{\mu}&=&\frac{1}{2}[u^{\dagger}\partial_{\mu}u+u\partial_{\mu}u^{\dagger}],\\
\chi_{\pm}&=&u^{\dagger}\chi u^{\dagger}\pm u\chi^{\dagger}u.
\end{eqnarray}
The leading order Lagrangian for the doubly heavy baryons is
\begin{eqnarray}
\mathcal{L}_{{\cal B}\phi}^{(1)}&=&\bar{{\mathcal B}}iv\cdot D{\cal B}-\bar{{\cal B}}^{*\mu}(iv\cdot D-\delta){\cal B}_{\mu}^{*}+2g_{1}\bar{{\cal B}}(S\cdot u){\cal B}\nonumber\\
&~~&+2g_{2}\bar{{\cal B}}^{*\mu}(u\cdot S){\cal B}_{\mu}^{*}+g_{3}(\bar{{\cal B}}^{*\mu}u_{\mu}{\cal B}+\bar{{\cal B}}u_{\mu}{\cal B}^{*\mu}).\label{op1}
\end{eqnarray}
where $\mathcal{B}$ and $\mathcal{B}^*$ are the doubly heavy baryon fields after heavy baryon reduction. The relativistic Lagrangian for doubly heavy baryon can be found in Ref.~\cite{Sun:2014aya}. In Eq.~(\ref{op1}), $v_\mu=(1,0,0,0)$ is the baryon velocity. $D_\mu=\partial_\mu+\Gamma_\mu$ is the chiral covariant derivative. $\delta=M^*-M$ is the mass splitting between spin-$1\over 2$ and spin-$3\over 2$ doubly charmed baryons. $S^\mu={i\over 2}\gamma_5 \sigma^{\mu\nu}v_{\nu}$ is the spin matrix. $g_{1,2,3}$ are the axial coupling constants. 

At the leading order, only the Weinberg-Tomozawa terms (the contact terms from the chiral connection) contribute to the scattering lengths through the tree diagram (a) in Fig.~\ref{tree}. The leading order vertices in the  tree diagrams (b) and (c) in Fig.~\ref{tree} arise from the axial coupling terms. These diagrams do not contribute to the scattering lengths at the leading order due to the vanishing structures $k\cdot S$, $k^{\mu}k^{\nu}P_{\mu\nu}$, and $\bar{\mathcal{B}}^{*\mu} k_{\mu}$ at threshold, where $P_{\mu\nu}=g_{\mu\nu}-v_{\mu}v_{\nu}+{4\over{d-1}}S_{\mu}S_{\nu}$ is the projection operator of Rarita-Schwinger field $\mathcal{B}^*_{\mu}$. $k^{\mu}=(m_{\phi},0)$ is the momentum of $\phi$ at threshold.

The $\mathcal{O}(p^2)$ Lagrangian contributing to the scattering lengths for the doubly charmed baryons reads: 
\begin{eqnarray}
{\cal L}_{{\cal B}\phi}^{(2)}&=&a_{0}\bar{{\cal B}}\text{Tr}[\chi_{+}]{\cal B}+a_{1}\bar{{\cal B}}\chi_{+}{\cal B}-a_{2}\bar{{\cal B}}\text{Tr}[(u\cdot v)^{2}]{\cal B}\nonumber \\
&~&-a_{3}\bar{{\cal B}}(u\cdot v)^{2}{\cal B}+a'_{0}\bar{{\cal B}}^{*\mu}\text{Tr}[\chi_{+}]{\cal B}_{\mu}^{*}+a'_{1}\bar{{\cal B}}^{*\mu}\chi_{+}{\cal B}_{\mu}^{*}\nonumber \\
&~&-a'_{2}\bar{{\cal B}}^{*\mu}\text{Tr}[(u\cdot v)^{2}]{\cal B}_{\mu}^{*}-a'_{3}\bar{{\cal B}}^{*\mu}(u\cdot v)^{2}{\cal B}_{\mu}^{*},\label{op2}
\end{eqnarray}
where $a_{1-4}$ and $a'_{1-4}$ are the LECs for the spin-$1\over 2$ and spin-$3\over 2$ doubly charmed baryons, respectively. At $\mathcal{O}(p^2)$, there are recoil terms derived from the leading order Lagrangian after heavy baryon expansion,
\begin{eqnarray}
{\cal L}_{{\cal B}\phi}^{(2,r.c)}=-\frac{g_{1}^{2}}{2M}\bar{{\cal B}}(v\cdot u)^{2}{\cal B}+\frac{g_{2}^{2}}{2M}\bar{{\cal B}}^{*\mu}(v\cdot u)^{2}{\cal B}_{\mu}^{*},\label{lag:rc2}
\end{eqnarray}
where the superscript ``$r.c$'' denote the recoil effect. The Lagrangians in Eqs.
\eqref{op2} and \eqref{lag:rc2} contribute the scattering lengths through tree diagram (a) in Fig.~\ref{tree}. At this order, one may also construct some $\mathcal{BB^*\phi}$ vertices
as in the discussion of the pion-nucleon interaction in Ref.~\cite{Fettes:2000bb}, which appear in the tree diagrams (b) and (c) in Fig.~\ref{tree}. For instance,
\begin{eqnarray}
\mathcal{L}^{(2)}_{\mathcal{BB^*}\phi}\sim i\bar{\mathcal{B}}^{*\mu}(\partial_{\mu}u_{\nu})v^{\nu}\mathcal{B}+\text{h.c.}.
\end{eqnarray}
However, these terms do not contribute to the $B_{cc}\phi$ and
$B_{cc}^*\phi$ scattering lengths due to the vanishing structures $\bar{\cal{B}}^{*\mu}k_{\mu}$ and $k_{\mu}k_{\nu}P^{\mu\nu}$.

At $\mathcal{O}(p^3)$, many terms contribute to the scattering lengths~\cite{Liu:2010bw}. In this work, we only keep the nonvanishing terms in the heavy diquark limit,
\begin{eqnarray}
{\cal L}_{{\cal B}\phi}^{(3)}=\tilde{\kappa}\bar{{\cal B}}[\chi_{-},v\cdot u]{\cal B}+\tilde{\kappa}'\bar{{\cal B}}^{*\mu}[\chi_{-},v\cdot u]{\cal B}_{\mu}^{*},
\end{eqnarray}
where $\tilde{\kappa}$ and $\tilde{\kappa}'$ are LECs. At this order, the recoil terms read,
\begin{eqnarray}
{\cal L_{{\cal B}\phi}}^{(3,r.c)}&&=\frac{g_{1}^{2}}{4M^{2}}\bar{{\cal B}}(v\cdot u)(iv\cdot D)(v\cdot u){\cal B}\nonumber \\
&-&\frac{g_{2}^{2}}{4M^{2}}\bar{{\cal B}}^{*\mu}(v\cdot u)(iv\cdot D+\delta)(v\cdot u){\cal B}_{\mu}^{*}.\label{lag:rc3}
\end{eqnarray}

\subsection{$\phi B_{cc}$ scattering}
There are eleven isospin-independent scattering lengths for the $\phi B_{cc}$. Some of them can be related to the others using the crossing symmetry. Only eight scattering lengths are independent. We list their analytical expressions of the threshold $T$-matrices order by order. The chiral orders are distinguished by the square bracket, which are labeled as superscripts. The analytical results without recoil effect read,
\begin{widetext}
\begin{align}
T_{\pi\Xi_{cc}}^{(3/2)} & =\left[-\frac{m_{\pi}}{2F_{\pi}^{2}}\right]^{(1)}+\left[-\frac{A_{1}m_{\pi}^{2}}{2F_{\pi}^{2}}\right]^{(2)}+\bigg[\frac{-4\tilde{\kappa}m_{\pi}^{3}}{F_{\pi}^{2}}-\frac{1}{16}V(m_{\pi}^{2},m_{\pi})-\frac{3}{16}V(m_{\pi}^{2},-m_{\pi})-\frac{1}{16}V(m_{K}^{2},-m_{\pi})\nonumber\\
 & ~~-m_{\pi}^{2}W_{1}(m_{\pi})+\frac{1}{9}m_{\pi}^{2}W_{1}(m_{\eta})\bigg]^{(3)},\nonumber\\
T_{\pi\Xi_{cc}}^{(1/2)} & =\left[\frac{m_{\pi}}{F_{\pi}^{2}}\right]^{(1)}+\left[-\frac{A_{1}m_{\pi}^{2}}{2F_{\pi}^{2}}\right]^{(2)}+\bigg[\frac{8\tilde{\kappa}m_{\pi}^{3}}{F_{\pi}^{2}}-\frac{1}{4}V(m_{\pi}^{2},m_{\pi})-\frac{3}{32}V(m_{K}^{2},m_{\pi})+\frac{1}{32}V(m_{K}^{2},-m_{\pi})\nonumber\\
 &~~ -m_{\pi}^{2}W_{1}(m_{\pi})+\frac{1}{9}m_{\pi}^{2}W_{1}(m_{\eta})\bigg]^{(3)},\nonumber\\
T_{\pi\Omega_{cc}}^{(1)} & =\left[0\right]^{(1)}+\left[-\frac{(A_{1}+A_{0})m_{\pi}^{2}}{4F_{\pi}^{2}}\right]^{(2)}+\left[0-\frac{1}{16}V(m_{K}^{2},m_{\pi})-\frac{1}{16}V(m_{K}^{2},-m_{\pi})+\frac{4}{9}m_{\pi}^{2}W_{1}(m_{\eta})\right]^{(3)},\nonumber\\
T_{K\Xi_{cc}}^{(1)} & =\left[-\frac{m_{K}}{2F_{K}^{2}}\right]^{(1)}+\left[-\frac{A_{1}m_{K}^{2}}{2F_{K}^{2}}\right]^{(2)}+\bigg[\frac{-4\tilde{\kappa}m_{K}^{3}}{F_{K}^{2}}-\frac{1}{16}V(m_{K}^{2},m_{K})-\frac{1}{32}V(m_{\pi}^{2},-m_{K})-\frac{1}{8}V(m_{K}^{2},-m_{K})\nonumber\\
 &~~~ -\frac{3}{32}V(m_{\eta}^{2},-m_{K})-\frac{2}{9}m_{K}^{2}W_{1}(m_{\eta})-\frac{2}{3}m_{K}^{2}U_{1}\bigg]^{(3)},\nonumber\\
T_{K\Xi_{cc}}^{(0)} & =\left[\frac{m_{K}}{2F_{K}^{2}}\right]^{(1)}+\left[-\frac{A_{0}m_{K}^{2}}{2F_{K}^{2}}\right]^{(2)}+\bigg[\frac{4\tilde{\kappa}m_{K}^{3}}{F_{K}^{2}}-\frac{1}{16}V(m_{K}^{2},m_{K})-\frac{3}{32}V(m_{\pi}^{2},-m_{K})+\frac{1}{8}V(m_{K}^{2},-m_{K})\nonumber\\
 &~~~ +\frac{3}{32}V(m_{\eta}^{2},-m_{K})-\frac{2}{9}m_{K}^{2}W_{1}(m_{\eta})+2m_{K}^{2}U_{1}\bigg]^{(3)},\nonumber\\
T_{K\Omega_{cc}}^{(1/2)} & =\left[\frac{m_{K}}{2F_{K}^{2}}\right]^{(1)}+\left[-\frac{A_{1}m_{K}^{2}}{2F_{K}^{2}}\right]^{(2)}+\bigg[\frac{4\tilde{\kappa}m_{K}^{3}}{F_{K}^{2}}-\frac{3}{32}V(m_{\pi}^{2},m_{K})-\frac{1}{16}V(m_{K}^{2},m_{K})-\frac{3}{32}V(m_{\eta}^{2},m_{K})\nonumber\\
 & ~~~-\frac{1}{16}V(m_{K}^{2},-m_{K})-\frac{8}{9}m_{K}^{2}W_{1}(m_{\eta})\bigg]^{(3)},\nonumber\\
T_{\eta\Xi_{cc}}^{(1/2)} & =\left[0\right]^{(1)}+\left[-\frac{(2A_{1}+A_{0})m_{\eta}^{2}}{6F_{\eta}^{2}}+\frac{2a_{1}(m_{\eta}^{2}-m_{\pi}^{2})}{3F_{\eta}^{2}}\right]^{(2)}+\bigg[0-\frac{3}{32}V(m_{K}^{2},m_{\eta})-\frac{3}{32}V(m_{K}^{2},-m_{\eta})\nonumber\\
 & ~~~+m_{\pi}^{2}W_{1}(m_{\pi})-\frac{4}{3}m_{K}^{2}W_{1}(m_{K})-\frac{1}{9}m_{\pi}^{2}W_{1}(m_{\eta})+\frac{4}{9}m_{\eta}^{2}W_{1}(m_{\eta})\bigg]^{(3)},\nonumber\\
T_{\eta\Omega_{cc}}^{(0)} & =\left[0\right]^{(1)}+\left[-\frac{(7A_{1}-A_{0})m_{\eta}^{2}}{12F_{\eta}^{2}}-\frac{4a_{1}(m_{\eta}^{2}-m_{\pi}^{2})}{3F_{\eta}^{2}}\right]^{(2)}+\bigg[0-\frac{3}{16}V(m_{K}^{2},m_{\eta})-\frac{3}{16}V(m_{K}^{2},-m_{\eta})\nonumber\\
 &~~~ -\frac{8}{3}m_{K}^{2}W_{1}(m_{K})-\frac{4}{9}m_{\pi}^{2}W_{1}(m_{\eta})+\frac{16}{9}m_{\eta}^{2}W_{1}(m_{\eta})\bigg]^{(3)},\label{Tbcc}
\end{align}
\end{widetext}
where the functions $V$, $W_1$ and $U_1$ are defined in Appendix~\ref{app:integrals}. In order to get the concise expressions, we have used the Gell-Mann-Okubo relation, $m_\eta^2=(4m_K^2-m_\pi^2)/3$. We define the $A_0$ and $A_1$ as
\begin{eqnarray}
&A_{1}	\equiv8a_{0}+4a_{1}+2a_{2}+a_{3},\nonumber\\
&A_{0}	\equiv8a_{0}-4a_{1}+2a_{2}-a_{3}.
\end{eqnarray}
There are six independent LECs in Eq.~(\ref{Tbcc}), $g_{1}$, $g_3$, $A_0$, $A_1$, $a_1$ and $\tilde{\kappa}$ to be determined.

The contribution from the recoil terms in Eqs.~(\ref{lag:rc2}) and (\ref{lag:rc3}) reads,
\begin{align}
T_{\pi\Xi_{cc}}^{(3/2,r.c)} & =\left[0\right]^{(1)}+\left[-\frac{g_{1}^{2}m_{\pi}^{2}}{4MF_{\pi}^{2}}\right]^{(2)}+\left[\frac{g_{1}^{2}m_{\pi}^{3}}{8M^{2}F_{\pi}^{2}}\right]^{(3)},\nonumber \\
T_{\pi\Xi_{cc}}^{(1/2,r.c)} & =\left[0\right]^{(1)}+\left[-\frac{g_{1}^{2}m_{\pi}^{2}}{4MF_{\pi}^{2}}\right]^{(2)}+\left[-\frac{g_{1}^{2}m_{\pi}^{3}}{4M^{2}F_{\pi}^{2}}\right]^{(3)},\nonumber \\
T_{\pi\Omega_{cc}}^{(1,r.c)} & =\left[0\right]^{(1)}+\left[0\right]^{(2)}+[0]^{(3)},\nonumber \\
T_{K\Xi_{cc}}^{(1,r.c)} & =\left[0\right]^{(1)}+\left[-\frac{g_{1}^{2}m_{K}^{2}}{4MF_{K}^{2}}\right]^{(2)}+\left[\frac{g_{1}^{2}m_{K}^{3}}{8M^{2}F_{K}^{2}}\right]^{(3)},\nonumber \\
T_{K\Xi_{cc}}^{(0,r.c)} & =\left[0\right]^{(1)}+\left[\frac{g_{1}^{2}m_{K}^{2}}{4MF_{K}^{2}}\right]^{(2)}+\left[-\frac{g_{1}^{2}m_{K}^{3}}{8M^{2}F_{K}^{2}}\right]^{(3)},\nonumber \\
T_{K\Omega_{cc}}^{(1/2,r.c)} & =\left[0\right]^{(1)}+\left[-\frac{g_{1}^{2}m_{K}^{2}}{4MF_{K}^{2}}\right]^{(2)}+\left[-\frac{g_{1}^{2}m_{K}^{3}}{8M^{2}F_{K}^{2}}\right]^{(3)},\nonumber \\
T_{\eta\Xi_{cc}}^{(1/2,r.c)} & =\left[0\right]^{(1)}+\left[-\frac{g_{1}^{2}m_{\eta}^{2}}{12MF_{\eta}^{2}}\right]^{(2)}+[0]^{(3)},\nonumber \\
T_{\eta\Omega_{cc}}^{(0,r.c)} & =\left[0\right]^{(1)}+\left[-\frac{g_{1}^{2}m_{\eta}^{2}}{3MF_{\eta}^{2}}\right]^{(2)}+[0]^{(3)}.
\end{align}

With the eight independent $T$-matrices, the others can be obtained through crossing symmetry,\begin{align}
T_{\bar{K},\Omega_{cc}}^{(1/2)}	&=[T_{K,\Omega_{cc}}^{(1/2)}]_{m_{K}\rightarrow-m_{K}},\nonumber\\
T_{\bar{K}\Xi_{cc}}^{(1)}	&=\frac{1}{2}[T_{K\Xi_{cc}}^{(1)}+T_{K\Xi_{cc}}^{(0)}]{}_{m_{K}\rightarrow-m_{K}},\nonumber \\
T_{\bar{K}\Xi_{cc}}^{(0)}	&=\frac{1}{2}[3T_{K\Xi_{cc}}^{(1)}-T_{K\Xi_{cc}}^{(0)}]{}_{m_{K}\rightarrow-m_{K}}.
\end{align}

\subsection{$\phi B^*_{cc}$ scattering}
The calculation of the $\phi B^*_{cc}$ scattering lengths is very similar. The $T$-matrices without the recoil effect are obtained by making the replacements as follows,
\begin{eqnarray}
&W_{1}\rightarrow W_{2},\quad U_{1}\rightarrow U_{2},\quad\tilde{\kappa}\rightarrow-\tilde{\kappa}',\nonumber\\
&a_{i}\rightarrow-a'_{i},\quad A_i\rightarrow A'_i,
\end{eqnarray}
with 
\begin{eqnarray}
&A'_{1}	\equiv 8a'_{0}+4a'_{1}+2a'_{2}+a'_{3},\nonumber\\
&A'_{0}	\equiv 8a'_{0}-4a'_{1}+2a'_{2}-a'_{3}.
\end{eqnarray}
The contribution of the recoil effect is obtained by making the replacements,
\begin{equation}
g_1^2\rightarrow g_2^2,\quad M\rightarrow M^*.
\end{equation}

\section{Chiral Lagrangian with heavy diquark-antiquark symmetry}\label{sec:symmetry}
In the analytical expressions of the $T$-matrices in Sec.~\ref{sec:Tmatrix}, there are eleven unknown LECs, $g_{1-3}$, $A_{0,1}$, $A'_{0,1}$, $a_1$, $a'_1$, $\tilde{\kappa}$ and $\tilde{\kappa}'$. In the heavy diquark limit, the LECs associated with $B_{cc}$ can be related to those of $B^*_{cc}$. Moreover, the HDA symmetry can relate the LECs associated with doubly charmed baryons to those of $D^{(*)}$ mesons, which have been determined in Refs.~\cite{Liu:2009uz,Liu:2011mi}.

In the following, we extend the formalism of the heavy quark symmetry~\cite{Falk:1991nq} to the doubly heavy systems. The doubly charmed baryons in the heavy diquark limit can be expressed as
\begin{equation}
\psi_{cc}^{\mu}\sim u_l A_h^{\mu},
\end{equation}
where $u_l$ is the spinor of the light quark. $A^{\mu}_h$ is the heavy diquark. The light and heavy components satisfy that
\begin{equation}
\slashed{v}u_l=u_l,\quad v\cdot A=0.
\end{equation}
We can construct the superfield and its conjugation,
\begin{eqnarray}
\psi^{\mu}_{cc}&=&{\cal B}^{*\mu}+\sqrt{\frac{1}{3}}(\gamma^{\mu}+v^{\mu})\gamma^{5}{\cal B},\nonumber \\
\bar{\psi}^{\mu}_{cc}&=&\bar{{\cal B}}^{*\mu}-\sqrt{\frac{1}{3}}\bar{{\cal B}}\gamma^{5}(\gamma^{\mu}+v^{\mu})
\end{eqnarray}
The construction details are given in Appendix~\ref{app:super}. One notices that the superfield for doubly charmed baryons with the heavy diquark symmetry have the same form as that of singly charmed sextet baryons~\cite{Falk:1991nq,Meng:2018gan}. However, they contain the different constituents. The superfield of singly heavy baryon can be represented as
\begin{equation}
\psi_{c}^{\mu}\sim u_hA_l^{\mu}.
\end{equation}
Thus, for the singly charmed baryons, the Lagrangian with heavy quark spin symmetry should have the form
\begin{equation}
\mathcal{L}\sim \bar{\psi}_c^{\mu}(\Gamma_h G_{l\mu\nu})\psi_c^{\nu}\sim (\bar{u}_h \Gamma_h u_h) (A_l^{*\mu}G_{l\mu\nu}A_l^{\nu}),
\end{equation}
where we divide the structure between the superfields into the $\Gamma_h$ in spinor space and $G_{l\mu\nu}$ in Lorentz space. The $\Gamma_h$ and $G_{l\mu\nu}$ are finally act on the heavy part and the light part in the superfield, respectively. Their subscripts denote the degrees of freedom they act on. The heavy quark symmetry constrains that $\Gamma_h$ should not flip the spin of the heavy quark. For the doubly charmed baryon, 
\begin{equation}
\mathcal{L}\sim \bar{\psi}_{cc}^{\mu}(\Gamma_l G_{h\mu\nu})\psi_{cc}^{\nu}\sim (\bar{u}_l \Gamma_l u_l) (A_h^{*\mu}G_{h\mu\nu}A_h^{\nu}).\label{lag:psicc}
\end{equation}
The structure in spinor space $\Gamma_l$ and the structure in Lorentz space $G_{h\mu\nu}$ will act on the light and heavy degrees of freedom, respectively. The heavy diquark symmetry has no constraint on $\Gamma_l$, but the $G_{h\mu\nu}$ should not change the polarization of the heavy diquark. Thus, the Lagrangian with the form $\bar{\psi}_{cc}^{\mu} \psi_{cc\mu}\sim A_h^{*\mu}A_{h\mu}$ guarantees the heavy diquark symmetry. With the superfield $\psi^{\mu}$ (we omit the subscript ``$cc$'' in the following), we can construct the Lagrangian with the heavy diquark symmetry and relate the LECs for spin-$1\over 2$ doubly charmed baryons to those for spin-$3\over 2$ ones. 

Since the doubly charmed baryons $(ccq)$ and singly charmed antimesons $(\bar{c}q)$ have the same light degree of freedom, they respond to the chiral transformation in the same way. With the HDA symmetry, their LECs can be related to each other. 

We notice that the superfield $H$ was introduced in Refs.~\cite{Falk:1991nq,Liu:2009uz,Liu:2011mi} to denote the $D$ and $D^*$ mesons in the heavy quark limit,
\begin{eqnarray}
H	&=&\frac{1+\slashed{v}}{2}\left(P_{\mu}^{*}\gamma^{\mu}+iP\gamma_{5}\right),\nonumber \\
\bar{H}	&=&\gamma^{0}H^{\dagger}\gamma^{0}=\left(P_{\mu}^{*\dagger}\gamma^{\mu}+iP^{\dagger}\gamma_{5}\right)\frac{1+\slashed{v}}{2},\nonumber \\
P&=&(D^0,D^+,D_s^+),\quad P_{\mu}^*=(D^{*0},D^{*+}，D_s^{*+})_{\mu}.
\end{eqnarray}
In the heavy quark limit, the field $P$ or $P^*_{\mu}$ only contains the operator annihilating a meson with $c$ quark. They can not create an antimeson with $\bar{c}$ quark. In order to describe the $(\bar{c}q)$ type antimesons, the $\tilde{H}$ is defined by~\cite{Grinstein:1992qt}
\begin{eqnarray}
\tilde{H}=C[\mathcal{C}H\mathcal{C}^{-1}]^TC^{-1},
\end{eqnarray}
where $C=i\gamma^2\gamma^0$. The details about charge conjugation of the field operators and building blocks are given in Appendix~\ref{app:super}. Thus the explicit form of $\tilde{H}$ and its conjugation read
\begin{eqnarray}
\tilde{H}&=&(\tilde{P}_{\mu}\gamma^{\mu}+i\tilde{P}\gamma_{5})\frac{1-\slashed{v}}{2},\nonumber \\
\bar{\tilde{H}}	&=&\frac{1-\slashed{v}}{2}(\tilde{P}_{\mu}^{*\dagger}\gamma^{\mu}+i\tilde{P}^{\dagger}\gamma_{5}),\nonumber \\
\tilde{P}&=&(\bar{D}^0,D^-,D^-_s)^T,\quad \tilde{P}_{\mu}=(\bar{D}^{*0},D^{*-},D^{*-}_s)^T.
\end{eqnarray}
The $H$ and $\tilde{H}$ can be represented by their quark components as
\begin{equation}
H\sim u_{h}\bar{v}_{l},\quad \tilde{H} \sim u_{l}\bar{v}_{h}.
\end{equation}
The general Lagrangian for the $\tilde{H}$ can be written as
\begin{equation}
{\cal L}\sim\bar{u}_{l}\Gamma_{l}u_{l}\bar{v}_{h}\Gamma_{h}v_{h}=\langle v_{h}\bar{u}_{l}\Gamma_{l}u_{l}\bar{v}_{h}\Gamma_{h}\rangle\sim\langle\bar{\tilde{H}}\Gamma_{l}\tilde{H}\Gamma_{h}\rangle,\label{lag:htilde}
\end{equation}
where $\langle...\rangle$ denotes the trace of the spinor indices. The heavy quark spin symmetry has constraint on the $\Gamma_h$ but not on the $\Gamma_l$. 

In the heavy quark (diquark) symmetry, the heavy quark (diquark) can be regarded as the spectator. Thus, $G_{h\mu\nu}=g_{\mu\nu}$ in Eq.~(\ref{lag:psicc}) and $\Gamma_h=\mathbf{1}$ in Eq.~(\ref{lag:htilde}).
In the HDA symmetry, the light degrees of freedom in Eqs.~(\ref{lag:psicc}) and (\ref{lag:htilde}) have the same dynamics. Thus, the $\Gamma_l$s in Eqs.~(\ref{lag:psicc}) and (\ref{lag:htilde}) are the same. Here we choose the proper normalization of fields to make the Lagrangians for doubly charmed baryons and charmed antimesons share the same LECs. We write the charge conjugation of heavy meson Lagrangians in Refs.~\cite{Liu:2009uz,Liu:2011mi} first. The Lagrangians of doubly charmed baryons in the HDA symmetry are followed,
\begin{eqnarray}
{\cal L}_{\tilde{H}\phi}^{(1)}	&=&-\langle\bar{\tilde{H}}iv\cdot D\tilde{H}\rangle+g\langle\bar{\tilde{H}}u_{\mu}\gamma^{\mu}\gamma_{5}\tilde{H}\rangle,\\
{\cal L}_{\psi\phi}^{(1)}	&=&-\bar{\psi}^{\mu}iv\cdot D\psi_{\mu}+g\bar{\psi}^{\mu}u\cdot\gamma\gamma_{5}\psi_{\mu};\\
{\cal L}_{\tilde{H}\phi}^{(2)}&=&c_{0}\langle\bar{\tilde{H}}\tilde{H}\rangle\text{Tr}(\chi_{+})+c_{1}\langle\bar{\tilde{H}}\chi_{+}\tilde{H}\rangle\nonumber \\
&-&c_{2}\langle\bar{\tilde{H}}\tilde{H}\rangle\text{Tr}(v\cdot uv\cdot u)-c_{3}\langle\bar{\tilde{H}}v\cdot uv\cdot u\tilde{H}\rangle,\\
{\cal L}_{\psi\phi}^{(2)}&=&c_{0}\bar{\psi}^{\mu}\psi_{\mu}\text{Tr}(\chi_{+})+c_{1}\bar{\psi}^{\mu}\chi_{+}\psi_{\mu}\nonumber \\
&-&c_{2}\bar{\psi}^{\mu}\psi_{\mu}\text{Tr}(v\cdot uv\cdot u)-c_{3}\bar{\psi}^{\mu}v\cdot uv\cdot u\psi_{\mu};\\
{\cal L}_{\tilde{H}\phi}^{(3)}&=&-\kappa\bar{\tilde{H}}[\chi_{-},v\cdot u]\tilde{H},\\
{\cal L}_{\psi\phi}^{(3)}&=&-\kappa\bar{\psi}^{\mu}[\chi_{-},v\cdot u]\psi_{\mu}.
\end{eqnarray}
Comparing these Lagrangians with those in Eqs.~(\ref{op1}) and (\ref{op2}), we get the relations among these LECs,
\begin{eqnarray}
&g_1={1\over 3}g,\quad g_2=g, \quad g_3=2\sqrt{1\over 3} g,\quad a_i=-a'_i=-c_i,\nonumber \\
&A_i=-A'_i=-C_i,\quad \tilde{\kappa}=-\tilde{\kappa}'=\kappa. \label{LECsrelation}
\end{eqnarray}
where $C_i$ are defined as,
\begin{eqnarray}
&C_{1}	\equiv8c_{0}+4c_{1}+2c_{2}+c_{3},\nonumber\\
&C_{0}	\equiv8c_{0}-4c_{1}+2c_{2}-c_{3}.
\end{eqnarray}

\section{Numerical results and discussions}\label{sec:num}
With the heavy quark symmetry, heavy diquark symmetry and HDA symmetry, the scattering lengths of the $\phi B^{(*)}_{cc}$ and $\phi D^{(*)}$ have very simple relations. In the heavy quark symmetry, the masses splitting between $D$ and $D^{*}$ vanishes. In the heavy diquark limit, the $B_{cc}$ and $B^*_{cc}$ are also degenerate states and recoil effect vanishes. In these limits, we compare our analytical results with those in Refs.~\cite{Liu:2009uz,Liu:2011mi},
\begin{eqnarray}
&T_{\phi B_{cc}}=T_{\phi B^*_{cc}}={1\over 2} T_{\phi \bar{D}}={1\over 2} T_{\phi \bar{D}^*},\\
&T_{\bar{\phi} D}=T_{\phi \bar{D}},\quad T_{\bar{\phi} D^*}=T_{\phi \bar{D}^*},\label{cparity}
\end{eqnarray}
where the relations in Eq.~(\ref{cparity}) are manifestation of the charge conjugation invariance. Their scattering lengths are related as,
\begin{equation}
a_{\phi B_{cc}}=a_{\phi B^*_{cc}}= a_{\bar{\phi} D}= a_{\bar{\phi} D^*}.\label{arelation}
\end{equation}
With the HDA symmetry, the heavy quark in the heavy meson and the heavy diquark in the doubly charmed baryon are both spectators. The light pseudoscalar mesons only interact with the light quark. Thus, their scattering lengths as physical observables are the same.

In the calculation of scattering lengths of $\phi D$, the $DD\phi$ vertices are forbidden since the parity and angular momentum conservation. Thus, the loop diagrams with intermediate $D$ mesons vanish. However, in the calculation of other scattering lengths, the doublet in the heavy quark (diquark) limit should both be considered as the intermediate states.

In the following, we will use the LECs of the $\phi D^{(*)}$ scattering in Refs.~\cite{Liu:2009uz,Liu:2011mi} to perform numerical analysis. The hadrons masses and the decay constants involved are as follows~\cite{Ebert:2002ig,Tanabashi:2018oca},
\begin{eqnarray}
&m_{\pi}	=139\text{ MeV},\quad m_{K}=494\text{ MeV},\quad m_{\eta}=\sqrt{\frac{4m_{K}^{2}-m_{\pi}^{2}}{3}},\nonumber \\
&F_{\pi}	=92\text{ MeV},\quad F_{K}=113\text{ MeV},\quad F_{\eta}=1.2F_{K},\nonumber \\
& m_{\Xi_{cc}}	=3621\text{MeV,}\quad m_{\Omega_{cc}}=3727\text{ MeV},\quad \delta=150\text{ MeV}\nonumber \\
& m_{\Xi_{cc}^{*}}	=m_{\Xi_{cc}}+\delta,\quad m_{\Omega_{cc}^{*}}=m_{\Omega_{cc}}+\delta.\label{paravalue}
\end{eqnarray}
Since only the $\Xi_{cc}^{++}$ has been observed in experiment, the masses of other doubly charmed baryons are obtained from the relativistic quark model~\cite{Ebert:2002ig}. We use the $\lambda=4\pi F_{\pi}$. The $c_1$ is obtained from the mass splitting between heavy mesons,
\begin{equation}
c_1={1\over 16}{M^2_{D_s}-M_D^2+M_{D_s^*}^2-M_{D^*}^2 \over M(m_K^2-m_{\pi}^2)}=0.12\text{ GeV}^{-1}.
\end{equation}
The axial coupling constants $g=0.59$ was determined from the decay width of $D^{*+}$~\cite{Ahmed:2001xc}. The $g_{1-3}$ can be related to the $g$. 

With the $g$ coupling constant, we can calculate the numerical results of the recoil effect, which are given in Table~\ref{num:recoil}. One can see that the recoil effect is extremely small. In the following calculation, we neglect the recoil effect. The mass splitting $\delta$ only occurs at the $\mathcal{O}(p^3)$ loop diagrams. In Table~\ref{delta}, we present the numerical results of the $\mathcal{O}(p^3)$ loop diagrams for $\delta=150\text{ MeV}$ and $\delta=0 \text{ MeV}$, respectively. One can see that the effect of the mass splitting is not significant. In the following calculation, we fix $\delta=150$ MeV.

\begin{table*}[htp]
\caption{The recoil correction to the $T$-matrices. The superscript denotes the isospin of this channel. $\mathcal{R} = |r.c/Re(T)|$, where $r.c$ is the recoil effect.  }\label{num:recoil}
\begin{tabular}{lcccccccc}
\toprule[1pt]\toprule[1pt]
Recoil effect & ~~~$T_{\pi\Xi_{cc}}^{(3/2)}$ & $T_{\pi\Xi_{cc}}^{(1/2)}$ & $T_{\pi\Omega_{cc}}^{(1)}$ & $T_{K\Xi_{cc}}^{(1)}$ & $T_{K\Xi_{cc}}^{(0)}$ & $T_{K\Omega_{cc}}^{(1/2)}$ & $T_{\eta\Xi_{cc}}^{(1/2)}$ & $T_{\eta\Omega_{cc}}^{(0)}$\tabularnewline
\midrule[1pt]
${\cal O}(p^{2})$ $(\times10^{-2}\text{fm})$ & -0.1 & -0.1 & 0.0 & -1.0 & 1.0 & -1.0 & 0.3 & -1.2\tabularnewline
 $\mathcal{R}_{\mathcal{O}(p^{2})}~(\%)$  &0.08 & 0.02 & 0.00 & 0.28 & 0.08 & 0.06 & 
 0.06 & 0.15 \tabularnewline
 \midrule[1pt]
${\cal O}(p^{3})(\times10^{-5}\text{fm})$ & 2.3 & -4.6 & 0.0 & 68.5 & 68.5 & 68.5 & 0.0 & 0.0\tabularnewline
 $\mathcal{R}_{\mathcal{O}(p^{3})}~(\%)$ & $<$0.01& $<$0.01 & 0.00 & 0.09 & 0.03 & 0.02 & 0.00 & 0.00 \tabularnewline
\bottomrule[1pt]\bottomrule[1pt]\\
\end{tabular}
\end{table*}

\begin{table*}[htp]
\caption{The numerical results of $\mathcal{O}(p^3)$ loop diagrams for $\delta=150\text{ MeV}$ and $\delta=0 \text{ MeV}$, respectively.}\label{delta}
\begin{tabular}{lcc|lcc}
\toprule[1pt]\toprule[1pt]
 & $\delta=150\text{ MeV}$ & $\delta=0\text{ MeV}$ &  & $\delta=150\text{ MeV}$ & $\delta=0\text{ MeV}$\tabularnewline
 \midrule[1pt]
$T_{\pi\Xi_{cc}}^{(3/2)}$ & $-0.47$ & $-0.49$ & $T_{\pi\Xi_{cc}^{*}}^{(3/2)}$ & $-0.53-0.01i$ & $-0.49$\tabularnewline
$T_{\pi\Xi_{cc}}^{(1/2)}$ & $0.31$ & $0.30$ & $T_{\pi\Xi_{cc}^{*}}^{(1/2)}$ & $0.25-0.01i$ & $0.30$\tabularnewline
$T_{\pi\Omega_{cc}}^{(1)}$ & $-0.56$ & $-0.56$ & $T_{\pi\Omega_{cc}^{*}}^{(1)}$ & $-0.56$ & $-0.56$\tabularnewline
$T_{K\Xi_{cc}}^{(1)}$ & $-3.08$ & $-3.08$ & $T_{K\Xi_{cc}^{*}}^{(1)}$ & $-3.11-0.003i$ & $-3.08$\tabularnewline
$T_{K\Xi_{cc}}^{(0)}$ & $3.74$ & $3.75$ & $T_{K\Xi_{cc}^{*}}^{(0)}$ & $3.84+0.001i$ & $3.75$\tabularnewline
$T_{K\Omega_{cc}}^{(1/2)}$ & $1.46+4.16i$ & $1.49+4.16i$ & $T_{K\Omega_{cc}^{*}}^{(1/2)}$ & $1.48+4.16i$ & $1.49+4.16i$\tabularnewline
$T_{\eta\Xi_{cc}}^{(1/2)}$ & $0.23+1.51i$ & $0.24+1.51i$ & $T_{\eta\Xi_{cc}^{*}}^{(1/2)}$ & $0.24+1.52i$ & $0.24+1.51i$\tabularnewline
$T_{\eta\Omega_{cc}}^{(0)}$ & $0.03+3.03i$ & $0.02+3.03i$ & $T_{\eta\Omega_{cc}^{*}}^{(0)}$ & $0.003+3.03i$ & $0.02+3.03i$\tabularnewline
$T_{\bar{K}\Xi_{cc}}^{(1)}$ & $-0.77+2.78i$ & $-0.76+2.77i$ & $T_{\bar{K}\Xi_{cc}^{*}}^{(1)}$ & $-0.73+2.77i$ & $-0.76+2.77i$\tabularnewline
$T_{\bar{K}\Xi_{cc}}^{(0)}$ & $4.48$ & $4.48$ & $T_{\bar{K}\Xi_{cc}^{*}}^{(0)}$ & $4.38-0.001i$ & $4.48$\tabularnewline
$T_{\bar{K}\Omega_{cc}}^{(1/2)}$ & $-2.39$ & $-2.36$ & $T_{\bar{K}\Omega_{cc}^{*}}^{(1/2)}$ & $-2.37$ & $-2.36$\tabularnewline
\bottomrule[1pt]\bottomrule[1pt]
\end{tabular}
\end{table*}
\subsection{Scenario I}
We use two scenarios to determine the other LECs. In the first scenario, three unknown LECs, $C_0$, $C_1$ and $\kappa$ are determined by fitting the lattice QCD results of the $\phi D$ scattering lengths. In Ref.~\cite{Liu:2011mi}, the authors adopted the preliminary lattice QCD simulation results~\cite{Liu:2008rza}. We update the fitting with the new lattice QCD results~\cite{Liu:2012zya}. In Ref.~\cite{Liu:2012zya}, the authors gave the scattering lengths of
	five channels. We choose three of them, $a_{\pi D}^{(3/2)}=-0.10\text{ fm}$,
	$a_{\pi D_s}^{(1)}=-0.002\text{ fm}$ and $a_{KD_s}^{(1/2)}=-0.18\text{ fm}$
	as input. The others can be used to estimate the uncertainty of our calculation. We get
\begin{eqnarray}
&C_{0}=0.81(30)\text{GeV}^{-1}, C_{1}=3.98(20)\text{GeV}^{-1},\nonumber \\ &\kappa=0.55(7)\text{GeV}^{-2}.\label{LES:SI}
\end{eqnarray}
The updated results of the $\phi D$ scattering are given in Table~\ref{Dsl}. The $T$-matrices and scattering lengths of the $\phi B_{cc}$ and $\phi B^*_{cc}$ are listed in Table~\ref{SI-Bcc}. The errors in Eq.~(\ref{LES:SI}) and Tables~\ref{Dsl} and \ref{SI-Bcc}
	arise from the uncertainties of our inputs. We vary the inputs
	in the range of their uncertainties and obtain the errors of
	numerical results. The errors related to truncating the perturbative expansion are not included.

	Our results in scenario I show good convergence of the chiral expansion
	for the channels in which only pions and $D$ or $\Xi_{cc}^{*}$ are involved. For the channels involving $\eta$, though the leading order contributions vanish, the next-to-leading order and the next-to-next-to leading order results also show good chiral convergence. In the
	channels involving the strange particles the convergence tends to be bad. The chiral
	symmetry is an approximate symmetry when the light quark mass goes to zero.
	For the $u$ and $d$ quarks, the chiral symmetry is a good approximation
	since their masses are very small. The $s$ quark mass is about 100 MeV, which
	will worsen the chiral convergence. To some extent, the worse chiral convergence
	for the channels involving the strange particles is not strongly related to
	the uncertainties of the unknown LECs. We take the $\phi D$ scatterings as an example.
	There are no unknown LECs in the leading order $T$-matrices and the LECs of
	the $\mathcal{O}(p^3)$ loop diagrams are all determined by fitting the experimental
	data. As shown in Table~\ref{Dsl}, the $\mathcal{O}(p^3)$ loop contribution
	is much smaller than the leading order T-matrices for the $\pi D$ channels.
	However, in some channels with strange particles, the $\mathcal{O}(p^3)$ loop
	contributions are either comparative with or even larger than the leading order results.

\begin{table*}[htp]
\caption{The $T$-matrices and scattering lengths of the $\phi D$ scattering (in units of fm), which are updated results in Ref.~\cite{Liu:2009uz}. The superscript ``$\dagger$'' denotes the input from the lattice QCD results.}\label{Dsl}
\begin{tabular}{lcccccccc}
	\toprule[1pt]\toprule[1pt]
	 & ~~~${\cal O}(p^{1})$~~~ & ~~~${\cal O}(p^{2})$~~~ & ${\cal O}(p^{3})$ loop & ${\cal O}(p^{3})$ tree & ${\cal O}(p^{3})$~~~ & ~~~Total~~~ & Scattering length & ~~lattice QCD~\cite{Liu:2012zya}\tabularnewline
\midrule[1pt]
$T_{\pi D}^{(3/2)}$ & $-3.2$ & $1.8$ & $-1.0$ & $-0.3$& $-1.3$ & $-2.7$ & $-0.100(2)$ & $-0.100(2)^{\dagger}$\tabularnewline
$T_{\pi D}^{(1/2)}$ & $6.4$ & $1.8$ & $0.6$ & $0.6$ & $1.1$ & $9.4$ & $0.35(1)$ & \tabularnewline
$T_{\pi D_{s}}^{(1)}$ & $0.0$ & $1.1$ & $-1.1$ & $0.0$ & $-1.1$ & $-0.1$ & $-0.002(1)$ & $-0.002(1)^{\dagger}$\tabularnewline
$T_{\bar{K}D}^{(1)}$ & $-7.6$ & $15.0$ & $-6.2$ & $-8.2$ & $-14.4$ & $-7.0$ & $-0.22(1)$ & $-0.20(1)$\tabularnewline
$T_{\bar{K}D}^{(0)}$ & $7.6$ & $3.0$ & $7.5$ & $8.2$ & $15.7$ & $26.4$ & $0.84(2)$ & $0.84(15)$\tabularnewline
$T_{\bar{K}D_{s}}^{(1/2)}$ & $7.6$ & $15.0$ & $3.0+8.3i$ & $8.2$ & $11.2+8.3i$ & $33.7+8.3i$ & $1.09(6)+0.27i$ & \tabularnewline
$T_{\eta D}^{(1/2)}$ & $0$ & $9.5$ & $0.5+3.0i$ & $0.0$ & $0.5+3.0i$ & $9.9+3.0i$ & $0.31(1)+0.09i$ & \tabularnewline
$T_{\eta D_{s}}^{(0)}$ & $0$ & $16.4$ & $0.04+6.1i$ & $0.0$ & $0.04+6.1i$ & $16.4+6.1i$ & $0.52(3)+0.19$ & \tabularnewline
$T_{KD}^{(1)}$ & $0$ & $9.0$ & $-1.5+5.5i$ & $0.0$ & $-1.5+5.5i$ & $7.5+5.5i$ & $0.24(1)+0.18i$ & \tabularnewline
$T_{KD}^{(0)}$ & $15.2$ & $20.9$ & $9.0$ & $16.4$ & $25.4$ & $61.5$ & $1.96(11)$ & \tabularnewline
$T_{KD_{s}}^{(1/2)}$ & $-7.6$ & $15.0$ & $-4.7$ & $-8.2$ & $-12.9$ & $-5.6$ & $-0.18(1)$ & $-0.18(1)^{\dagger}$\tabularnewline
\bottomrule[1pt]\bottomrule[1pt]
\end{tabular}
\end{table*}

\begin{table*}[htp]
\caption{The $T$-matrices and scattering lengths of $\phi B^{(*)}_{cc}$ scattering in scenario I (in units of fm). The LECs are determined as Eq.~(\ref{LES:SI}). }\label{SI-Bcc}
\begin{tabular}{lccccc}
\toprule[1pt]\toprule[1pt]
S-I & $~~~{\cal O}(p^{1})~~~$ & $~~~{\cal O}(p^{2})~~~$ & $~~~{\cal O}(p^{3})~~~$ & ~~~Total~~~ & Scattering Length\tabularnewline
\midrule[1pt]
$T_{\pi\Xi_{cc}}^{(3/2)}$ & $-1.61$ & $0.89$ & $-0.61$ & $-1.33$ & $-0.100(2)$\tabularnewline
$T_{\pi\Xi_{cc}}^{(1/2)}$ & $3.22$ & $0.89$ & $0.59$ & $4.71$ & $0.36(1)$\tabularnewline
$T_{\pi\Omega_{cc}}^{(1)}$ & $0.00$ & $0.54$ & $-0.56$ & $-0.02$ & $-0.002(1)$\tabularnewline
$T_{K\Xi_{cc}}^{(1)}$ & $-3.81$ & $7.48$ & $-7.18$ & $-3.52$ & $-0.25(1)$\tabularnewline
$T_{K\Xi_{cc}}^{(0)}$ & $3.81$ & $1.52$ & $7.84$ & $13.17$ & $0.92(2)$\tabularnewline
$T_{K\Omega_{cc}}^{(1/2)}$ & $3.81$ & $7.48$ & $5.55+4.16i$ & $16.84+4.16i$ & $1.18(6)+0.29i$\tabularnewline
$T_{\eta\Xi_{cc}}^{(1/2)}$ & $0.0$ & $4.73$ & $0.23+1.51i$ & $4.96+1.51i$ & $0.34(1)+0.10i$\tabularnewline
$T_{\eta\Omega_{cc}}^{(0)}$ & $0.0$ & $8.20$ & $0.03+3.03i$ & $8.23+3.03i$ & $0.57(3)+0.21i$\tabularnewline
$T_{\bar{K}\Xi_{cc}}^{(1)}$ & $0.00$ & $4.50$ & $-0.77+2.77i$ & $3.73+2.77i$ & $0.26(1)+0.19i$\tabularnewline
$T_{\bar{K}\Xi_{cc}}^{(0)}$ & $7.62$ & $10.45$ & $12.68$ & $30.75$ & $2.15(11)$\tabularnewline
$T_{\bar{K}\Omega_{cc}}^{(1/2)}$ & $-3.81$ & $7.48$ & $-6.49$ & $-2.83$ & $-0.20(1)$\tabularnewline
\midrule[1pt]
$T_{\pi\Xi_{cc}^{*}}^{(3/2)}$ & $-1.61$ & $0.89$ & $-0.67-0.01i$ & $-1.39-0.01i$ & $-0.11(2)-0.001i$\tabularnewline
$T_{\pi\Xi_{cc}^{*}}^{(1/2)}$ & $3.22$ & $0.89$ & $0.53-0.01i$ & $4.65-0.01i$ & $0.36(1)-0.001i$\tabularnewline
$T_{\pi\Omega_{cc}^{*}}^{(1)}$ & $0.00$ & $0.54$ & $-0.56$ & $-0.03$ & $-0.002(1)$\tabularnewline
$T_{K\Xi_{cc}^{*}}^{(1)}$ & $-3.81$ & $7.48$ & $-7.21-0.0003i$ & $-3.54-0.0003i$ & $-0.25(1)-2\times10^{-5}i$\tabularnewline
$T_{K\Xi_{cc}^{*}}^{(0)}$ & $3.81$ & $1.52$ & $7.94+0.001i$ & $13.28+0.001i$ & $0.93(2)+7\times10^{-5}i$\tabularnewline
$T_{K\Omega_{cc}^{*}}^{(1/2)}$ & $3.81$ & $7.48$ & $5.58+4.16i$ & $16.87+4.16i$ & $1.19(6)+0.29i$\tabularnewline
$T_{\eta\Xi_{cc}^{*}}^{(1/2)}$ & $0.0$ & $4.73$ & $0.24+1.52i$ & $4.96+1.52i$ & $0.34(1)+0.11i$\tabularnewline
$T_{\eta\Omega_{cc}^{*}}^{(0)}$ & $0.0$ & $8.20$ & $0.003+3.03i$ & $8.20+3.03i$ & $0.57(3)+0.21i$\tabularnewline
$T_{\bar{K}\Xi_{cc}^{*}}^{(1)}$ & $0.00$ & $4.50$ & $-0.73+2.77i$ & $3.77+2.77i$ & $0.27(1)+0.20i$\tabularnewline
$T_{\bar{K}\Xi_{cc}^{*}}^{(0)}$ & $7.62$ & $10.45$ & $12.58-0.001i$ & $30.65-0.001i$ & $2.16(11)-7\times10^{-5}i$\tabularnewline
$T_{\bar{K}\Omega_{cc}^{*}}^{(1/2)}$ & $-3.81$ & $7.48$ & $-6.46$ & $-2.80$ & $-0.20(1)$\tabularnewline
\bottomrule[1pt]\bottomrule[1pt] 
\end{tabular}
\end{table*} 
\subsection{Scenario II}
In Ref.~\cite{Liu:2011mi}, the $\phi D^*$ scattering lengths are investigated. The LECs are obtained using the resonance saturation model. The resonances such as the scalar singlet $\sigma(600)$, scalar octet, $\kappa(800)$, $a_0(980)$, $f_0(980)$ and the $D_{s1}(2460)$ are used to estimate the $\phi\phi D^*D^*$ vertex at threshold. In the second scenario, we use these LECs in Ref.~\cite{Liu:2011mi} directly,
\begin{eqnarray}
&c_0=0.10\text{GeV}^{-1},\quad c_1=0.12\text{GeV}^{-1},\quad c_2=-0.30\text{GeV}^{-1},\nonumber \\
&c_3=0.42\text{GeV}^{-1},\quad \kappa=-0.33 \text{GeV}^{-2}.\label{LEC:SII}
\end{eqnarray}
The $T$-matrices and scattering lengths are listed in Table~\ref{SII-Bcc}. The resonance saturation model will bring
	the uncertainties to LECs. In Ref.~\cite{Liu:2011mi}, the authors show
	that the uncertainties of their LECs are from 20\% to 60\%.
	In this work, we assume all LECs have the uncertainty 40\%
	and estimate the errors of scattering length in Table~\ref{SII-Bcc}. The errors related to truncating the perturbative expansion are not included.
	
We notice that the chiral convergence in scenario II is good for most of the channels. The LECs we obtained in scenario II is more natural than those in scenario I, which will be discussed in the next subsection.

\subsection{Discussion}
In both scenarios, the  $\phi B_{cc}$ scattering lengths are almost the same as those for $\phi B^*_{cc}$ channels. Their difference only arises from the mass splitting in the loop diagram, which has been shown to be small. In the most channels, the two scenarios give the similar results, at least with the same sign, except the $a_{\bar{K}\Xi_{cc}^{(*)}}^{(1)}$. The leading result for the $[\bar{K}\Xi^{(*)}_{cc}]^{(1)}$ channel vanishes. The uncertainty of LECs at the next-to-leading order gives rise to the different results in two scenarios. This issue can be clarified if more precise experiment or lattice QCD results appear.

In our convention, the positive (negative) sign of the scattering length indicates the interaction for this channel is attractive (repulsive). One notice that the interactions for $[\pi\Xi^{(*)}_{cc}]^{(1/2)}$, $[K\Xi^{(*)}_{cc}]^{(0)}$, $[K\Omega^{(*)}_{cc}]^{(1/2)}$, $[\eta\Xi^{(*)}_{cc}]^{(1/2)}$, $[\eta\Omega^{(*)}_{cc}]^{(0)}$ and $[\bar{K}\Xi^{(*)}_{cc}]^{(0)}$ channels are attractive. Bound states may appear in these channels. The $[\bar{K}\Xi^{(*)}_{cc}]^{(0)}$ channels are very interesting. The interaction in the two channels are the most attractive. With the HDA symmetry, the $[\bar{K}\Xi^{(*)}_{cc}]^{(0)}$ are the partner channels of $[K D^{(*)}]^{(0)}$, which are associated with the $D_{s0}^*(2317)$ ($D_{s1}(2460)$). The partner state of $D_{s0}^*(2317)$ may appear as a bound states in the $[\bar{K}\Xi_{cc}]^{(0)}$ channel. The couple channel effect from the $[\bar{K}\Xi_{cc}^*]^{(0)}$ may help to form the partner state of $D_{s1}(2460)$.

 In Ref.~\cite{Guo:2017vcf}, the author adopted the chiral unitary approach for the $\phi B_{cc}$ system. One bound states below the $[\bar{K}\Xi_{cc}]^{(0)}$ threshold and two resonant structures in the $[\pi\Xi_{cc}]^{(1/2)}$, $[K\Omega_{cc}]^{(1/2)}$ and  $[\eta\Xi_{cc}]^{(1/2)}$ coupled-channel scattering were found, which are consistent with our calculation.

 In the ChPT, the amplitude is expanded as,
	\begin{equation}
	\mathcal{M}=\mathcal{M}^{(0)}\sum_{\nu }\left({q\over \Lambda_\chi}\right)^{\nu} \mathcal{F}(g_i)
	\end{equation}
	where $\mathcal{M}^{(0)}$ is the leading order amplitude.
	$\mathcal{F}$ is a function of LECs. Thus the convergence of the
	chiral expansion is based on an implicit assumption that
	$\mathcal{F}(g_i)$ is of order one. This assumption is the so-called
	naturalness assumption. In order to judge the naturalness of our
	LECs, we define the $\alpha$ as
	\begin{eqnarray}
	|T^{(\nu)}|=\frac{m_{\phi}}{2F_{\phi}^{2}}\left(\frac{m_{\phi}}{4\pi F_{\phi}}\right)^{\nu-1}\alpha^{(\nu)}.
	\end{eqnarray}
	where $\frac{m_{\phi}}{2F_{\phi}^{2}}$ is the approximation of the
	leading order $T$-matrices. The LECs we determined in scenario I and
	scenario II only contribute to the $\mathcal{O}(p^2)$  and
	$\mathcal{O}(p^3)$ tree diagrams. We present the $\alpha^{(2)}$ and
	$\alpha^{(3)}_{tree}$ in Table~\ref{natrualness}. Most of the $\alpha$s are of
	order 1. The $\alpha$s in scenario II, are closer to 1 than those in
	scenario I. Thus the results in scenario II show better convergence.
	In the scenario I, most of the $\alpha$s are larger than 3. Since
	${m_{\pi} \over 4\pi F_{\phi}}\approx 0.12$, thus the channels with
	pions still have the good convergence. Since ${m_K\over 4\pi
		F_{\phi}}\approx {m_{\eta}\over 4\pi F_{\phi}}\approx 0.3$, the LECs
	in scenarios I is not natural enough to ensure the good convergence
	of the chiral expansion in channels with $K$ and $\eta$ mesons. In
	our calculation, we perform the chiral expansion to the
	next-to-next-to leading order. When the bound state, virtual state
	or resonance appear, the finite order chiral expansion will fail.
	The bad convergence in scenario I might also stem from the presence
	of these non-perturbative phenomena. One way to solve the problem is
	to unitarize the amplitude to resume the
	high order corrections~\cite{Guo:2009ct,Guo:2018tjx}.

\begin{table*}
\caption{The $T$-matrices and scattering lengths of $\phi B^{(*)}_{cc}$ scattering in scenario II (in units of fm). The LECs are determined as Eq.~(\ref{LEC:SII}).}\label{SII-Bcc}
\begin{tabular}{lccccc}
\toprule[1pt]\toprule[1pt]
S-II & ~~~${\cal O}(p^{1})$~~~ & ~~~${\cal O}(p^{2})$~~~ & ~~~${\cal O}(p^{3})$~~~ & ~~~Total~~~ & Scattering length\tabularnewline
\midrule[1pt] 
$T_{\pi\Xi_{cc}}^{(3/2)}$ & $-1.61$ & $0.22$ & $-0.39$ & $-1.78$ & $-0.13(2)$\tabularnewline
$T_{\pi\Xi_{cc}}^{(1/2)}$ & $3.22$ & $0.22$ & $0.15$ & $3.59$ & $0.27(2)$\tabularnewline
$T_{\pi\Omega_{cc}}^{(1)}$ & $0.00$ & $0.03$ & $-0.56$ & $-0.53$ & $-0.04(1)$\tabularnewline
$T_{K\Xi_{cc}}^{(1)}$ & $-3.81$ & $1.84$ & $-0.63$ & $-2.60$ & $-0.18(15)$\tabularnewline
$T_{K\Xi_{cc}}^{(0)}$ & $3.81$ & $-1.32$ & $1.29$ & $3.78$ & $0.26(17)$\tabularnewline
$T_{K\Omega_{cc}}^{(1/2)}$ & $3.81$ & $1.84$ & $-1.00+4.16i$ & $4.66+4.16i$ & $0.33(16)+0.29i$\tabularnewline
$T_{\eta\Xi_{cc}}^{(1/2)}$ & $0.0$ & $0.50$ & $0.23+1.51i$ & $0.73+1.51i$ & $0.05(7)+0.10i$\tabularnewline
$T_{\eta\Omega_{cc}}^{(0)}$ & $0.0$ & $2.58$ & $0.03+3.03i$ & $2.61+3.03i$ & $0.18(13)+0.21i$\tabularnewline
$T_{\bar{K}\Xi_{cc}}^{(1)}$ & $0.00$ & $0.26$ & $-0.77+2.77i$ & $-0.50+2.77i$ & $-0.03(7)+0.19i$\tabularnewline
$T_{\bar{K}\Xi_{cc}}^{(0)}$ & $7.62$ & $3.42$ & $-0.42$ & $10.62$ & $0.74(26)$\tabularnewline
$T_{\bar{K}\Omega_{cc}}^{(1/2)}$ & $-3.81$ & $1.84$ & $0.06$ & $-1.91$ & $-0.13(16)$\tabularnewline
\midrule[1pt]
$T_{\pi\Xi_{cc}^{*}}^{(3/2)}$ & $-1.61$ & $0.22$ & $-0.45-0.01i$ & $-1.84-0.01i$ & $0.14(2)-0.001i$\tabularnewline
$T_{\pi\Xi_{cc}^{*}}^{(1/2)}$ & $3.22$ & $0.22$ & $0.09-0.01i$ & $3.53-0.01i$ & $0.27(2)-0.001i$\tabularnewline
$T_{\pi\Omega_{cc}^{*}}^{(1)}$ & $0.00$ & $0.03$ & $-0.56$ & $-0.53$ & $-0.04(1)$\tabularnewline
$T_{K\Xi_{cc}^{*}}^{(1)}$ & $-3.81$ & $1.84$ & $-0.66-0.0003i$ & $-2.63-0.0003i$ & $-0.18(15)-2\times10^{-5}i$\tabularnewline
$T_{K\Xi_{cc}^{*}}^{(0)}$ & $3.81$ & $-1.32$ & $1.40+0.001i$ & $3.89+0.001i$ & $0.27(17)+7\times10^{-5}i$\tabularnewline
$T_{K\Omega_{cc}^{*}}^{(1/2)}$ & $3.81$ & $1.84$ & $-0.97+4.16i$ & $4.68+4.16i$ & $0.33(16)+0.29i$\tabularnewline
$T_{\eta\Xi_{cc}^{*}}^{(1/2)}$ & $0.0$ & $0.50$ & $0.24+1.52i$ & $0.74+1.52i$ & $0.05(7)+0.11i$\tabularnewline
$T_{\eta\Omega_{cc}^{*}}^{(0)}$ & $0.0$ & $2.58$ & $0.003+3.03i$ & $2.58+3.03i$ & $0.18(13)+0.21i$\tabularnewline
$T_{\bar{K}\Xi_{cc}^{*}}^{(1)}$ & $0.00$ & $0.26$ & $-0.73+2.77i$ & $-0.46+2.77i$ & $-0.03(7)+0.20i$\tabularnewline
$T_{\bar{K}\Xi_{cc}^{*}}^{(0)}$ & $7.62$ & $3.42$ & $-0.52-0.001i$ & $10.52-0.001i$ & $0.74(26)-7\times10^{-5}i$\tabularnewline
$T_{\bar{K}\Omega_{cc}^{*}}^{(1/2)}$ & $-3.81$ & $1.84$ & $0.09$ & $-1.88$ & $-0.13(16)$\tabularnewline
\bottomrule[1pt]\bottomrule[1pt]
\end{tabular}
\end{table*}

The HDA symmetry we adopted is an approximation,
which will bring uncertainties to our results for the $\phi
B_{cc}^{(*)}$ scattering. The validity of the HDA symmetry in the
charm sector was discussed in Ref.~\cite{Cohen:2006jg}. The author showed that the
charm quark is not heavy enough to justify the pointlike nature of
the double heavy diquark which is necessary for the HDA symmetry.
However, the author also claimed that one can not completely rule
out the possibility that HDA could hold approximately. The author
constructed a phenomenological quark model with coulombic and
confining potential, in which the size of the heavy diquark is small
enough to render the approximate HDA symmetry. In this work, we
adopt the HDA symmetry as an assumption to relate the LECs in the
$\phi B_{cc}^{(*)}$ scattering to those in the $\phi \bar{D}^{(*)}$
scattering. Our results and predictions also provide a way to test
the HDA symmetry.

\begin{table*}
	\caption{The naturalness of the LECs we determined. }\label{natrualness}
	\begin{tabular}{ccccccccccccc}
		\toprule[1pt]\toprule[1pt]
		&  & $[\pi\Xi_{cc}]^{({3\over 2})}$ & $[\pi\Xi_{cc}]^{({1\over 2})}$ & $[\pi\Omega_{cc}]^{(1)}$ & $[K\Xi_{cc}]^{(1)}$ & $[K\Xi_{cc}]^{(0)}$ & $[K\Omega_{cc}]^{({1\over 2})}$ & $[\eta\Xi_{cc}]^{({1\over 2})}$ & $[\eta\Omega_{cc}]^{(0)}$ & $[\bar{K}\Xi_{cc}]^{(1)}$ & $[\bar{K}\Xi_{cc}]^{(0)}$ & $[\bar{K}\Omega_{cc}]^{({1\over 2})}$\tabularnewline
		\midrule[1pt]
		\multirow{2}{*}{S-I} & $\alpha^{(2)}$ & 4.62 & 4.62 & 2.78 & 4.62 & 0.94 & 4.62 & 3.39 & 5.22 & 2.78 & 0.90 & 4.62\tabularnewline
		& $\alpha_{tree}^{(3)}$ & 5.95 & 11.91 & 0 & 5.95 & 5.95 & 5.95 & 0 & 0 & 0 & 5.95 & 5.95\tabularnewline
		\midrule[1pt]
		\multirow{2}{*}{S-II} & $\alpha^{(2)}$ & 2.67 & 2.67 & 1.63 & 2.67 & 0.58 & 2.67 & 1.97 & 3.02 & 1.63 & 0.46 & 2.68\tabularnewline
		& $\alpha_{tree}^{(3)}$ & 3.56 & 7.12 & 0 & 3.56 & 3.56 & 3.56 & 0 & 0 & 0 & 3.56 & 3.56\tabularnewline
		\bottomrule[1pt]\bottomrule[1pt]
	\end{tabular}
\end{table*}

\section{Conclusion}\label{sec:concl}
In this work, we adopt the heavy baryon chiral perturbation theory (HBChPT) to calculate the scattering lengths of the $\phi B_{cc}^{(*)}$ to the $\mathcal{O}(p^3)$. The analytical expressions are presented with the mass splitting between $B_{cc}$ and $B_{cc}^*$ and the recoil effect. In order to obtain the numerical results, we use the heavy diquark-antiquark (HDA) symmetry to relate the $B_{cc}^{(*)}$ to $D^{(*)}$. With the HDA symmetry, we construct the Lagrangians with the help of superfields. We use the LECs of $\phi D^{(*)}$ scattering in Refs.~\cite{Liu:2009uz} and \cite{Liu:2011mi} as two scenarios, respectively. The LECs of the $\phi D$ scattering in \cite{Liu:2009uz} are obtained by fitting the lattice QCD results. In the first scenario, we update the $\phi D$ scattering lengths with the new lattice QCD input and give the numerical results in doubly charmed sector. In this scenario, the chiral convergence for some channels is not good enough. Two reasons may account for the convergence. First, the large strange quark mass will worsen the chiral convergence for channels with strange particles. Second, the presence of the bound states, virtual states and resonances may destroy the convergence of the chiral expansion. In the second scenario, we use the LECs estimated with resonance saturation model in Ref.~\cite{Liu:2011mi}. The chiral convergence in this scenario is quite reasonable.

We calculate the recoil effect and the mass splitting effect numerically. These two effects are less important. Our final numerical results of scattering lengths in two scenarios are consistent with each other. The interactions for the $[\pi\Xi^{(*)}_{cc}]^{(1/2)}$, $[K\Xi^{(*)}_{cc}]^{(0)}$, $[K\Omega^{(*)}_{cc}]^{(1/2)}$, $[\eta\Xi^{(*)}_{cc}]^{(1/2)}$, $[\eta\Omega^{(*)}_{cc}]^{(0)}$ and $[\bar{K}\Xi^{(*)}_{cc}]^{(0)}$ channels are attractive. The bound states might appear in these channels, which can be searched for in the future experiments. The $[\bar{K}\Xi^{(*)}_{cc}]^{(0)}$ channel is the most attractive, which may help to form the partner states of the $D_{s0}^*(2317)$ ($D_{s1}(2460)$) in the doubly heavy sector.

 A by-product of this work is our contraction of the chiral Lagrangians with HDA symmetry. Unlike the Lagrangians in Ref.~\cite{Hu:2005gf}, we adopt the four component spinors in our construction, which ensure the Lagrangians formally covariant. It is very convenient to extend our approach to other much more complicated chiral Lagrangians with HDA symmetry.

\section*{ACKNOWLEDGMENTS}

L. Meng is very grateful to G. J. Wang, X. Z. Weng, X. L. Chen and W. Z. Deng for
very helpful discussions. This project is supported by the National
Natural Science Foundation of China under Grants 11575008,
11621131001 and 973 program.

\begin{appendix}

\section{Integrals}\label{app:integrals}
\begin{itemize}
\item Integral with one meson propagator:
\begin{eqnarray}
\Delta(m^{2})&=&i\int\frac{d^{d}\lambda^{4-d}}{(2\pi)^{d}}\frac{1}{l^{2}-m^{2}+i\epsilon}\nonumber \\
&=&2m^{2}\left(L(\lambda)+\frac{1}{32\pi^{2}}\text{ln}\frac{m^{2}}{\lambda^{2}}\right),
\end{eqnarray}
with the divergence term:
\begin{eqnarray}
L(\lambda)=\frac{\lambda^{d-4}}{16\pi^{2}}\left[\frac{1}{d-4}+\frac{1}{2}\left(\gamma_{E}-1-\text{ln}4\pi\right)\right],
\end{eqnarray}
where $d$ is the dimension.

\begin{widetext}
\item Integrals with one meson propagator and one baryon propagator:
\begin{eqnarray}
i\int\frac{d^{d}l\,\lambda^{4-d}}{(2\pi)^{d}}\frac{[1,~ l_{\alpha},~ l_{\alpha}l_{\beta}]}{(l^{2}-m^{2}+i\epsilon)(\omega+v\cdot l+i\epsilon)}=[J_{0}(m^{2},\omega),~v_{\alpha}J_{1}(m^{2},\omega),~ g_{\alpha\beta}J_{2}(m^{2},\omega)+v_{\alpha}v_{\beta}J_{3}(m^{2},\omega)],
\end{eqnarray}
\begin{eqnarray}
J_{0}(m^{2},\omega)=\begin{cases}
{\displaystyle \frac{-\omega}{8\pi^{2}}(1-\ln\frac{m^{2}}{\lambda^{2}})+\frac{\sqrt{\omega^{2}-m^{2}}}{4\pi^{2}}({\rm \text{arccosh}}\frac{\omega}{m}-i\pi)+4\omega L(\lambda)} & (\omega>m)\\
{\displaystyle \frac{-\omega}{8\pi^{2}}(1-\ln\frac{m^{2}}{\lambda^{2}})+\frac{\sqrt{m^{2}-\omega^{2}}}{4\pi^{2}}\arccos\frac{-\omega}{m}+4\omega L(\lambda)} & (\omega^{2}<m^{2})\\
{\displaystyle \frac{-\omega}{8\pi^{2}}(1-\ln\frac{m^{2}}{\lambda^{2}})-\frac{\sqrt{\omega^{2}-m^{2}}}{4\pi^{2}}{\rm \text{arccosh}}\frac{-\omega}{m}+4\omega L(\lambda)} & (\omega<-m)
\end{cases},
\end{eqnarray}
\begin{eqnarray}
J_{1}(m^{2},\omega)&	=&-\omega J_{0}(m^{2},\omega)+\Delta(m^{2}),\\
J_{2}(m^{2},\omega)	&=&\frac{1}{d-1}[(m^{2}-\omega^{2})J_{0}(m^{2},\omega)+\omega\Delta(m^{2})],\\
J_{3}(m^{2},\omega)	&=&-\omega J_{1}(m^{2},\omega)-J_{2}(m^{2},\omega)
\end{eqnarray}

\item Integrals with two meson propagators and one baryon propagator:
\begin{eqnarray}
&~~&i\int\frac{d^{d}l\,\lambda^{4-d}}{(2\pi)^{d}}\frac{[1,~l_{\alpha},~l_{\alpha}l_{\beta}]}{(l^{2}-m_{1}^{2}+i\epsilon)(l^{2}-m_{2}^{2}+i\epsilon)(\omega+v\cdot l+i\epsilon)} \nonumber \\
&=&[\Lambda_{0}(m_{1}^{2},m_{2}^{2},\omega),~v_{\alpha}\Lambda_{1}(m_{1}^{2},m_{2}^{2},\omega),~g_{\alpha\beta}\Lambda_{2}(m_{1}^{2},m_{2}^{2},\omega)+v_{\alpha}v_{\beta}\Lambda_{3}(m_{1}^{2},m_{2}^{2},\omega)],
\end{eqnarray}
\begin{eqnarray}
\Lambda_{i}(m_{1}^{2},m_{2}^{2},\omega)=\frac{1}{m_{1}^{2}-m_{2}^{2}}[J_{i}(m_{1}^{2},\omega)-J_{i}(m_{2}^{2},\omega)].
\end{eqnarray}
\item Functions used in this work:
\begin{align}
V(m^{2},\omega)&=\frac{4\omega^{2}J_{0}(m^{2},\omega)}{F^{4}},\nonumber\\
W_{1}(m)	&=-\frac{3g_{1}^{2}\Lambda_{2}(m,m,0)+2g_{3}^{2}\Lambda_{2}(m,m,-\delta)}{4F_{\phi}^{4}},\nonumber\\
W_{2}(m)	&=-\frac{5g_{2}^{2}\Lambda_{2}(m,m,0)+3g_{3}^{2}\Lambda_{2}(m,m,\delta)}{12F_{\phi}^{4}},\nonumber\\
U_{1}	&=-\frac{3g_{1}^{2}\Lambda_{2}(m_{\eta},m_{\pi},0)+2g_{3}^{2}\Lambda_{2}(m_{\eta},m_{\pi},-\delta)}{4F_{\phi}^{4}},\nonumber\\
U_{2}	&=-\frac{5g_{2}^{2}\Lambda_{2}(m_{\eta},m_{\pi},0)+3g_{3}^{2}\Lambda_{2}(m_{\eta},m_{\pi},\delta)}{12F_{\phi}^{4}}.
\end{align}
where $\delta$ is the mass splitting between $\Xi_{cc}(\Omega_{cc})$ and $\Xi_{cc}^*(\Omega_{cc}^*)$. Its value are given in Eq.~(\ref{paravalue}).
\end{widetext}
\end{itemize}

\section{Superfield}\label{app:super}
We extend the formalism of heavy quark symmetry~\cite{Falk:1991nq} to the doubly heavy system. The doubly charmed baryons in the heavy diquark limit can be constructed as
\begin{equation}
\psi_{cc}^{\mu}\sim u_l A_h^{\mu},
\end{equation}
where $u_l$ is the spinor of the light quark. $A^{\mu}_h$ is the heavy diquark. The light and heavy component satisfy that
\begin{equation}
\slashed{v}u_l=u_l,\quad v\cdot A=0.
\end{equation}
We can decompose the superfield into the spin-$3\over 2$ and spin-$1\over 2$ parts.
\begin{equation}
\psi_{cc}^{\mu}=\psi_{3/2}^{\mu}+\psi_{1/2}^{\mu},
\end{equation}
The general form of the spin-$3\over 2$ part can be parameterized as
\begin{equation}
\psi_{3/2}^{\mu}=\psi^{\mu}+bv^{\mu}\gamma\cdot\psi_{cc} +c\gamma^{\mu}\gamma\cdot\psi_{cc}.
\end{equation}
We use the properties of the spin-$3\over 2$ Rarita-Schwinger vector spinor $\psi_{3/2}^{\mu}$,
\begin{equation}
\slashed{v}\psi^{\mu}_{3/2}=\psi^{\mu}_{3/2},\quad v_{\mu} \psi^{\mu}_{3/2}=0,\quad \gamma_{\mu} \psi^{\mu}_{3/2}=0, \label{condition}
\end{equation}
and get 
\begin{equation}
b=c=-\frac{1}{3}, \quad \psi_{1/2}^{\mu}={1\over 3}(v^{\mu}+\gamma^{\mu})\gamma\cdot \psi_{cc}.
\end{equation}
The $\psi_{1/2}$ part can be rewritten using a spin-$1\over 2$ spinor $\tilde{\psi}$
\begin{equation}
\psi^\mu_{1/2}=\sqrt{1\over 3}(\gamma^{\mu}+v^{\mu})\gamma_5\tilde{\psi},\quad \text{with }\tilde{\psi}=\sqrt{1\over 3}\gamma_5\gamma_{\mu}\psi_{1/2}^{\mu},
\end{equation}
Thus, the superfield and its conjugation read
\begin{eqnarray}
\psi^{\mu}_{cc}&=&{\cal B}^{*\mu}+\sqrt{\frac{1}{3}}(\gamma^{\mu}+v^{\mu})\gamma^{5}{\cal B},\nonumber \\
\nonumber \\
\bar{\psi}^{\mu}_{cc}&=&\bar{{\cal B}}^{*\mu}-\sqrt{\frac{1}{3}}\bar{{\cal B}}\gamma^{5}(\gamma^{\mu}+v^{\mu}).
\end{eqnarray}

Under the charge conjugation, the building blocks and fields transform as 
\begin{eqnarray}
&\mathcal{C}u\mathcal{C}^{-1}=u^T,\quad \mathcal{C}\Gamma_{\mu}\mathcal{C}^{-1}=-\Gamma_{\mu}^T,\quad \mathcal{C}u_{\mu}\mathcal{C}^{-1}=u_{\mu}^T,\nonumber \\ 
&\mathcal{C}\chi_{\pm}\mathcal{C}^{-1}=\chi_{\pm}^T,\quad \mathcal{C}P\mathcal{C}^{-1}=\tilde{P},\quad \mathcal{C}P^*_{\mu}\mathcal{C}^{-1}=-\tilde{P}_{\mu}^*.\nonumber \\
\end{eqnarray}

\end{appendix}

\vfil \thispagestyle{empty}

\newpage
\bibliography{Bib}

\begin{thebibliography}{47}%
\makeatletter
\providecommand \@ifxundefined [1]{%
 \@ifx{#1\undefined}
}%
\providecommand \@ifnum [1]{%
 \ifnum #1\expandafter \@firstoftwo
 \else \expandafter \@secondoftwo
 \fi
}%
\providecommand \@ifx [1]{%
 \ifx #1\expandafter \@firstoftwo
 \else \expandafter \@secondoftwo
 \fi
}%
\providecommand \natexlab [1]{#1}%
\providecommand \enquote  [1]{``#1''}%
\providecommand \bibnamefont  [1]{#1}%
\providecommand \bibfnamefont [1]{#1}%
\providecommand \citenamefont [1]{#1}%
\providecommand \href@noop [0]{\@secondoftwo}%
\providecommand \href [0]{\begingroup \@sanitize@url \@href}%
\providecommand \@href[1]{\@@startlink{#1}\@@href}%
\providecommand \@@href[1]{\endgroup#1\@@endlink}%
\providecommand \@sanitize@url [0]{\catcode `\\12\catcode `\$12\catcode
  `\&12\catcode `\#12\catcode `\^12\catcode `\_12\catcode `\%12\relax}%
\providecommand \@@startlink[1]{}%
\providecommand \@@endlink[0]{}%
\providecommand \url  [0]{\begingroup\@sanitize@url \@url }%
\providecommand \@url [1]{\endgroup\@href {#1}{\urlprefix }}%
\providecommand \urlprefix  [0]{URL }%
\providecommand \Eprint [0]{\href }%
\providecommand \doibase [0]{http://dx.doi.org/}%
\providecommand \selectlanguage [0]{\@gobble}%
\providecommand \bibinfo  [0]{\@secondoftwo}%
\providecommand \bibfield  [0]{\@secondoftwo}%
\providecommand \translation [1]{[#1]}%
\providecommand \BibitemOpen [0]{}%
\providecommand \bibitemStop [0]{}%
\providecommand \bibitemNoStop [0]{.\EOS\space}%
\providecommand \EOS [0]{\spacefactor3000\relax}%
\providecommand \BibitemShut  [1]{\csname bibitem#1\endcsname}%
\let\auto@bib@innerbib\@empty
\bibitem [{\citenamefont {Aaij}\ \emph {et~al.}(2017)\citenamefont {Aaij} \emph
  {et~al.}}]{Aaij:2017ueg}%
  \BibitemOpen
  \bibfield  {author} {\bibinfo {author} {\bibfnamefont {R.}~\bibnamefont
  {Aaij}} \emph {et~al.} (\bibinfo {collaboration} {LHCb}),\ }\href {\doibase
  10.1103/PhysRevLett.119.112001} {\bibfield  {journal} {\bibinfo  {journal}
  {Phys. Rev. Lett.}\ }\textbf {\bibinfo {volume} {119}},\ \bibinfo {pages}
  {112001} (\bibinfo {year} {2017})},\ \Eprint
  {http://arxiv.org/abs/1707.01621} {arXiv:1707.01621 [hep-ex]} \BibitemShut
  {NoStop}%
\bibitem [{\citenamefont {Hyodo}\ \emph {et~al.}(2017)\citenamefont {Hyodo},
  \citenamefont {Liu}, \citenamefont {Oka},\ and\ \citenamefont
  {Yasui}}]{Hyodo:2017hue}%
  \BibitemOpen
  \bibfield  {author} {\bibinfo {author} {\bibfnamefont {T.}~\bibnamefont
  {Hyodo}}, \bibinfo {author} {\bibfnamefont {Y.-R.}\ \bibnamefont {Liu}},
  \bibinfo {author} {\bibfnamefont {M.}~\bibnamefont {Oka}}, \ and\ \bibinfo
  {author} {\bibfnamefont {S.}~\bibnamefont {Yasui}},\ }\href@noop {} {\
  (\bibinfo {year} {2017})},\ \Eprint {http://arxiv.org/abs/1708.05169}
  {arXiv:1708.05169 [hep-ph]} \BibitemShut {NoStop}%
\bibitem [{\citenamefont {Lü}\ \emph {et~al.}(2017)\citenamefont {Lü},
  \citenamefont {Wang}, \citenamefont {Xiao},\ and\ \citenamefont
  {Zhong}}]{Lu:2017meb}%
  \BibitemOpen
  \bibfield  {author} {\bibinfo {author} {\bibfnamefont {Q.-F.}\ \bibnamefont
  {Lü}}, \bibinfo {author} {\bibfnamefont {K.-L.}\ \bibnamefont {Wang}},
  \bibinfo {author} {\bibfnamefont {L.-Y.}\ \bibnamefont {Xiao}}, \ and\
  \bibinfo {author} {\bibfnamefont {X.-H.}\ \bibnamefont {Zhong}},\ }\href
  {\doibase 10.1103/PhysRevD.96.114006} {\bibfield  {journal} {\bibinfo
  {journal} {Phys. Rev.}\ }\textbf {\bibinfo {volume} {D96}},\ \bibinfo {pages}
  {114006} (\bibinfo {year} {2017})},\ \Eprint
  {http://arxiv.org/abs/1708.04468} {arXiv:1708.04468 [hep-ph]} \BibitemShut
  {NoStop}%
\bibitem [{\citenamefont {Yan}\ \emph {et~al.}(2018)\citenamefont {Yan},
  \citenamefont {Liu}, \citenamefont {Gonzàlez-Solís}, \citenamefont {Guo},
  \citenamefont {Hanhart}, \citenamefont {Meißner},\ and\ \citenamefont
  {Zou}}]{Yan:2018zdt}%
  \BibitemOpen
  \bibfield  {author} {\bibinfo {author} {\bibfnamefont {M.-J.}\ \bibnamefont
  {Yan}}, \bibinfo {author} {\bibfnamefont {X.-H.}\ \bibnamefont {Liu}},
  \bibinfo {author} {\bibfnamefont {S.}~\bibnamefont {Gonzàlez-Solís}},
  \bibinfo {author} {\bibfnamefont {F.-K.}\ \bibnamefont {Guo}}, \bibinfo
  {author} {\bibfnamefont {C.}~\bibnamefont {Hanhart}}, \bibinfo {author}
  {\bibfnamefont {U.-G.}\ \bibnamefont {Meißner}}, \ and\ \bibinfo {author}
  {\bibfnamefont {B.-S.}\ \bibnamefont {Zou}},\ }\href {\doibase
  10.1103/PhysRevD.98.091502} {\bibfield  {journal} {\bibinfo  {journal} {Phys.
  Rev.}\ }\textbf {\bibinfo {volume} {D98}},\ \bibinfo {pages} {091502}
  (\bibinfo {year} {2018})},\ \Eprint {http://arxiv.org/abs/1805.10972}
  {arXiv:1805.10972 [hep-ph]} \BibitemShut {NoStop}%
\bibitem [{\citenamefont {Weng}\ \emph {et~al.}(2018)\citenamefont {Weng},
  \citenamefont {Chen},\ and\ \citenamefont {Deng}}]{Weng:2018mmf}%
  \BibitemOpen
  \bibfield  {author} {\bibinfo {author} {\bibfnamefont {X.-Z.}\ \bibnamefont
  {Weng}}, \bibinfo {author} {\bibfnamefont {X.-L.}\ \bibnamefont {Chen}}, \
  and\ \bibinfo {author} {\bibfnamefont {W.-Z.}\ \bibnamefont {Deng}},\ }\href
  {\doibase 10.1103/PhysRevD.97.054008} {\bibfield  {journal} {\bibinfo
  {journal} {Phys. Rev.}\ }\textbf {\bibinfo {volume} {D97}},\ \bibinfo {pages}
  {054008} (\bibinfo {year} {2018})},\ \Eprint
  {http://arxiv.org/abs/1801.08644} {arXiv:1801.08644 [hep-ph]} \BibitemShut
  {NoStop}%
\bibitem [{\citenamefont {Wang}\ \emph
  {et~al.}(2017{\natexlab{a}})\citenamefont {Wang}, \citenamefont {Xing},\ and\
  \citenamefont {Xu}}]{Wang:2017azm}%
  \BibitemOpen
  \bibfield  {author} {\bibinfo {author} {\bibfnamefont {W.}~\bibnamefont
  {Wang}}, \bibinfo {author} {\bibfnamefont {Z.-P.}\ \bibnamefont {Xing}}, \
  and\ \bibinfo {author} {\bibfnamefont {J.}~\bibnamefont {Xu}},\ }\href
  {\doibase 10.1140/epjc/s10052-017-5363-y} {\bibfield  {journal} {\bibinfo
  {journal} {Eur. Phys. J.}\ }\textbf {\bibinfo {volume} {C77}},\ \bibinfo
  {pages} {800} (\bibinfo {year} {2017}{\natexlab{a}})},\ \Eprint
  {http://arxiv.org/abs/1707.06570} {arXiv:1707.06570 [hep-ph]} \BibitemShut
  {NoStop}%
\bibitem [{\citenamefont {Xiao}\ \emph {et~al.}(2017)\citenamefont {Xiao},
  \citenamefont {Wang}, \citenamefont {Lu}, \citenamefont {Zhong},\ and\
  \citenamefont {Zhu}}]{Xiao:2017udy}%
  \BibitemOpen
  \bibfield  {author} {\bibinfo {author} {\bibfnamefont {L.-Y.}\ \bibnamefont
  {Xiao}}, \bibinfo {author} {\bibfnamefont {K.-L.}\ \bibnamefont {Wang}},
  \bibinfo {author} {\bibfnamefont {Q.-f.}\ \bibnamefont {Lu}}, \bibinfo
  {author} {\bibfnamefont {X.-H.}\ \bibnamefont {Zhong}}, \ and\ \bibinfo
  {author} {\bibfnamefont {S.-L.}\ \bibnamefont {Zhu}},\ }\href {\doibase
  10.1103/PhysRevD.96.094005} {\bibfield  {journal} {\bibinfo  {journal} {Phys.
  Rev.}\ }\textbf {\bibinfo {volume} {D96}},\ \bibinfo {pages} {094005}
  (\bibinfo {year} {2017})},\ \Eprint {http://arxiv.org/abs/1708.04384}
  {arXiv:1708.04384 [hep-ph]} \BibitemShut {NoStop}%
\bibitem [{\citenamefont {Wang}\ \emph
  {et~al.}(2017{\natexlab{b}})\citenamefont {Wang}, \citenamefont {Yu},\ and\
  \citenamefont {Zhao}}]{Wang:2017mqp}%
  \BibitemOpen
  \bibfield  {author} {\bibinfo {author} {\bibfnamefont {W.}~\bibnamefont
  {Wang}}, \bibinfo {author} {\bibfnamefont {F.-S.}\ \bibnamefont {Yu}}, \ and\
  \bibinfo {author} {\bibfnamefont {Z.-X.}\ \bibnamefont {Zhao}},\ }\href
  {\doibase 10.1140/epjc/s10052-017-5360-1} {\bibfield  {journal} {\bibinfo
  {journal} {Eur. Phys. J.}\ }\textbf {\bibinfo {volume} {C77}},\ \bibinfo
  {pages} {781} (\bibinfo {year} {2017}{\natexlab{b}})},\ \Eprint
  {http://arxiv.org/abs/1707.02834} {arXiv:1707.02834 [hep-ph]} \BibitemShut
  {NoStop}%
\bibitem [{\citenamefont {Cheng}\ and\ \citenamefont
  {Shi}(2018)}]{Cheng:2018mwu}%
  \BibitemOpen
  \bibfield  {author} {\bibinfo {author} {\bibfnamefont {H.-Y.}\ \bibnamefont
  {Cheng}}\ and\ \bibinfo {author} {\bibfnamefont {Y.-L.}\ \bibnamefont
  {Shi}},\ }\href {\doibase 10.1103/PhysRevD.98.113005} {\bibfield  {journal}
  {\bibinfo  {journal} {Phys. Rev.}\ }\textbf {\bibinfo {volume} {D98}},\
  \bibinfo {pages} {113005} (\bibinfo {year} {2018})},\ \Eprint
  {http://arxiv.org/abs/1809.08102} {arXiv:1809.08102 [hep-ph]} \BibitemShut
  {NoStop}%
\bibitem [{\citenamefont {Li}\ \emph {et~al.}(2017)\citenamefont {Li},
  \citenamefont {Meng}, \citenamefont {Liu},\ and\ \citenamefont
  {Zhu}}]{Li:2017cfz}%
  \BibitemOpen
  \bibfield  {author} {\bibinfo {author} {\bibfnamefont {H.-S.}\ \bibnamefont
  {Li}}, \bibinfo {author} {\bibfnamefont {L.}~\bibnamefont {Meng}}, \bibinfo
  {author} {\bibfnamefont {Z.-W.}\ \bibnamefont {Liu}}, \ and\ \bibinfo
  {author} {\bibfnamefont {S.-L.}\ \bibnamefont {Zhu}},\ }\href {\doibase
  10.1103/PhysRevD.96.076011} {\bibfield  {journal} {\bibinfo  {journal} {Phys.
  Rev.}\ }\textbf {\bibinfo {volume} {D96}},\ \bibinfo {pages} {076011}
  (\bibinfo {year} {2017})},\ \Eprint {http://arxiv.org/abs/1707.02765}
  {arXiv:1707.02765 [hep-ph]} \BibitemShut {NoStop}%
\bibitem [{\citenamefont {Meng}\ \emph
  {et~al.}(2017{\natexlab{a}})\citenamefont {Meng}, \citenamefont {Li},
  \citenamefont {Liu},\ and\ \citenamefont {Zhu}}]{Meng:2017dni}%
  \BibitemOpen
  \bibfield  {author} {\bibinfo {author} {\bibfnamefont {L.}~\bibnamefont
  {Meng}}, \bibinfo {author} {\bibfnamefont {H.-S.}\ \bibnamefont {Li}},
  \bibinfo {author} {\bibfnamefont {Z.-W.}\ \bibnamefont {Liu}}, \ and\
  \bibinfo {author} {\bibfnamefont {S.-L.}\ \bibnamefont {Zhu}},\ }\href
  {\doibase 10.1140/epjc/s10052-017-5447-8} {\bibfield  {journal} {\bibinfo
  {journal} {Eur. Phys. J.}\ }\textbf {\bibinfo {volume} {C77}},\ \bibinfo
  {pages} {869} (\bibinfo {year} {2017}{\natexlab{a}})},\ \Eprint
  {http://arxiv.org/abs/1710.08283} {arXiv:1710.08283 [hep-ph]} \BibitemShut
  {NoStop}%
\bibitem [{\citenamefont {Li}\ \emph {et~al.}(2018)\citenamefont {Li},
  \citenamefont {Meng}, \citenamefont {Liu},\ and\ \citenamefont
  {Zhu}}]{Li:2017pxa}%
  \BibitemOpen
  \bibfield  {author} {\bibinfo {author} {\bibfnamefont {H.-S.}\ \bibnamefont
  {Li}}, \bibinfo {author} {\bibfnamefont {L.}~\bibnamefont {Meng}}, \bibinfo
  {author} {\bibfnamefont {Z.-W.}\ \bibnamefont {Liu}}, \ and\ \bibinfo
  {author} {\bibfnamefont {S.-L.}\ \bibnamefont {Zhu}},\ }\href {\doibase
  10.1016/j.physletb.2017.12.031} {\bibfield  {journal} {\bibinfo  {journal}
  {Phys. Lett.}\ }\textbf {\bibinfo {volume} {B777}},\ \bibinfo {pages} {169}
  (\bibinfo {year} {2018})},\ \Eprint {http://arxiv.org/abs/1708.03620}
  {arXiv:1708.03620 [hep-ph]} \BibitemShut {NoStop}%
\bibitem [{\citenamefont {Bahtiyar}\ \emph {et~al.}(2018)\citenamefont
  {Bahtiyar}, \citenamefont {Can}, \citenamefont {Erkol}, \citenamefont {Oka},\
  and\ \citenamefont {Takahashi}}]{Bahtiyar:2018vub}%
  \BibitemOpen
  \bibfield  {author} {\bibinfo {author} {\bibfnamefont {H.}~\bibnamefont
  {Bahtiyar}}, \bibinfo {author} {\bibfnamefont {K.~U.}\ \bibnamefont {Can}},
  \bibinfo {author} {\bibfnamefont {G.}~\bibnamefont {Erkol}}, \bibinfo
  {author} {\bibfnamefont {M.}~\bibnamefont {Oka}}, \ and\ \bibinfo {author}
  {\bibfnamefont {T.~T.}\ \bibnamefont {Takahashi}},\ }\href {\doibase
  10.1103/PhysRevD.98.114505} {\bibfield  {journal} {\bibinfo  {journal} {Phys.
  Rev.}\ }\textbf {\bibinfo {volume} {D98}},\ \bibinfo {pages} {114505}
  (\bibinfo {year} {2018})},\ \Eprint {http://arxiv.org/abs/1807.06795}
  {arXiv:1807.06795 [hep-lat]} \BibitemShut {NoStop}%
\bibitem [{\citenamefont {Hiller~Blin}\ \emph {et~al.}(2018)\citenamefont
  {Hiller~Blin}, \citenamefont {Sun},\ and\ \citenamefont
  {Vicente~Vacas}}]{Blin:2018pmj}%
  \BibitemOpen
  \bibfield  {author} {\bibinfo {author} {\bibfnamefont {A.~N.}\ \bibnamefont
  {Hiller~Blin}}, \bibinfo {author} {\bibfnamefont {Z.-F.}\ \bibnamefont
  {Sun}}, \ and\ \bibinfo {author} {\bibfnamefont {M.~J.}\ \bibnamefont
  {Vicente~Vacas}},\ }\href {\doibase 10.1103/PhysRevD.98.054025} {\bibfield
  {journal} {\bibinfo  {journal} {Phys. Rev.}\ }\textbf {\bibinfo {volume}
  {D98}},\ \bibinfo {pages} {054025} (\bibinfo {year} {2018})},\ \Eprint
  {http://arxiv.org/abs/1807.01059} {arXiv:1807.01059 [hep-ph]} \BibitemShut
  {NoStop}%
\bibitem [{\citenamefont {Meng}\ \emph
  {et~al.}(2017{\natexlab{b}})\citenamefont {Meng}, \citenamefont {Li},\ and\
  \citenamefont {Zhu}}]{Meng:2017fwb}%
  \BibitemOpen
  \bibfield  {author} {\bibinfo {author} {\bibfnamefont {L.}~\bibnamefont
  {Meng}}, \bibinfo {author} {\bibfnamefont {N.}~\bibnamefont {Li}}, \ and\
  \bibinfo {author} {\bibfnamefont {S.-L.}\ \bibnamefont {Zhu}},\ }\href
  {\doibase 10.1103/PhysRevD.95.114019} {\bibfield  {journal} {\bibinfo
  {journal} {Phys. Rev.}\ }\textbf {\bibinfo {volume} {D95}},\ \bibinfo {pages}
  {114019} (\bibinfo {year} {2017}{\natexlab{b}})},\ \Eprint
  {http://arxiv.org/abs/1704.01009} {arXiv:1704.01009 [hep-ph]} \BibitemShut
  {NoStop}%
\bibitem [{\citenamefont {Meng}\ \emph
  {et~al.}(2018{\natexlab{a}})\citenamefont {Meng}, \citenamefont {Li},\ and\
  \citenamefont {Zhu}}]{Meng:2017udf}%
  \BibitemOpen
  \bibfield  {author} {\bibinfo {author} {\bibfnamefont {L.}~\bibnamefont
  {Meng}}, \bibinfo {author} {\bibfnamefont {N.}~\bibnamefont {Li}}, \ and\
  \bibinfo {author} {\bibfnamefont {S.-l.}\ \bibnamefont {Zhu}},\ }\href
  {\doibase 10.1140/epja/i2018-12578-2} {\bibfield  {journal} {\bibinfo
  {journal} {Eur. Phys. J.}\ }\textbf {\bibinfo {volume} {A54}},\ \bibinfo
  {pages} {143} (\bibinfo {year} {2018}{\natexlab{a}})},\ \Eprint
  {http://arxiv.org/abs/1707.03598} {arXiv:1707.03598 [hep-ph]} \BibitemShut
  {NoStop}%
\bibitem [{\citenamefont {Chen}\ \emph {et~al.}(2018)\citenamefont {Chen},
  \citenamefont {Wang}, \citenamefont {Hosaka},\ and\ \citenamefont
  {Liu}}]{Chen:2018pzd}%
  \BibitemOpen
  \bibfield  {author} {\bibinfo {author} {\bibfnamefont {R.}~\bibnamefont
  {Chen}}, \bibinfo {author} {\bibfnamefont {F.-L.}\ \bibnamefont {Wang}},
  \bibinfo {author} {\bibfnamefont {A.}~\bibnamefont {Hosaka}}, \ and\ \bibinfo
  {author} {\bibfnamefont {X.}~\bibnamefont {Liu}},\ }\href {\doibase
  10.1103/PhysRevD.97.114011} {\bibfield  {journal} {\bibinfo  {journal} {Phys.
  Rev.}\ }\textbf {\bibinfo {volume} {D97}},\ \bibinfo {pages} {114011}
  (\bibinfo {year} {2018})},\ \Eprint {http://arxiv.org/abs/1804.02961}
  {arXiv:1804.02961 [hep-ph]} \BibitemShut {NoStop}%
\bibitem [{\citenamefont {Guo}(2017)}]{Guo:2017vcf}%
  \BibitemOpen
  \bibfield  {author} {\bibinfo {author} {\bibfnamefont {Z.-H.}\ \bibnamefont
  {Guo}},\ }\href {\doibase 10.1103/PhysRevD.96.074004} {\bibfield  {journal}
  {\bibinfo  {journal} {Phys. Rev.}\ }\textbf {\bibinfo {volume} {D96}},\
  \bibinfo {pages} {074004} (\bibinfo {year} {2017})},\ \Eprint
  {http://arxiv.org/abs/1708.04145} {arXiv:1708.04145 [hep-ph]} \BibitemShut
  {NoStop}%
\bibitem [{\citenamefont {Savage}\ and\ \citenamefont
  {Wise}(1990)}]{Savage:1990di}%
  \BibitemOpen
  \bibfield  {author} {\bibinfo {author} {\bibfnamefont {M.~J.}\ \bibnamefont
  {Savage}}\ and\ \bibinfo {author} {\bibfnamefont {M.~B.}\ \bibnamefont
  {Wise}},\ }\href {\doibase 10.1016/0370-2693(90)90035-5} {\bibfield
  {journal} {\bibinfo  {journal} {Phys. Lett.}\ }\textbf {\bibinfo {volume}
  {B248}},\ \bibinfo {pages} {177} (\bibinfo {year} {1990})}\BibitemShut
  {NoStop}%
\bibitem [{\citenamefont {Cohen}\ and\ \citenamefont
  {Hohler}(2006)}]{Cohen:2006jg}%
  \BibitemOpen
  \bibfield  {author} {\bibinfo {author} {\bibfnamefont {T.~D.}\ \bibnamefont
  {Cohen}}\ and\ \bibinfo {author} {\bibfnamefont {P.~M.}\ \bibnamefont
  {Hohler}},\ }\href {\doibase 10.1103/PhysRevD.74.094003} {\bibfield
  {journal} {\bibinfo  {journal} {Phys. Rev.}\ }\textbf {\bibinfo {volume}
  {D74}},\ \bibinfo {pages} {094003} (\bibinfo {year} {2006})},\ \Eprint
  {http://arxiv.org/abs/hep-ph/0606084} {arXiv:hep-ph/0606084 [hep-ph]}
  \BibitemShut {NoStop}%
\bibitem [{\citenamefont {Hu}\ and\ \citenamefont {Mehen}(2006)}]{Hu:2005gf}%
  \BibitemOpen
  \bibfield  {author} {\bibinfo {author} {\bibfnamefont {J.}~\bibnamefont
  {Hu}}\ and\ \bibinfo {author} {\bibfnamefont {T.}~\bibnamefont {Mehen}},\
  }\href {\doibase 10.1103/PhysRevD.73.054003} {\bibfield  {journal} {\bibinfo
  {journal} {Phys. Rev.}\ }\textbf {\bibinfo {volume} {D73}},\ \bibinfo {pages}
  {054003} (\bibinfo {year} {2006})},\ \Eprint
  {http://arxiv.org/abs/hep-ph/0511321} {arXiv:hep-ph/0511321 [hep-ph]}
  \BibitemShut {NoStop}%
\bibitem [{\citenamefont {Aubert}\ \emph {et~al.}(2003)\citenamefont {Aubert}
  \emph {et~al.}}]{Aubert:2003fg}%
  \BibitemOpen
  \bibfield  {author} {\bibinfo {author} {\bibfnamefont {B.}~\bibnamefont
  {Aubert}} \emph {et~al.} (\bibinfo {collaboration} {BaBar}),\ }\href
  {\doibase 10.1103/PhysRevLett.90.242001} {\bibfield  {journal} {\bibinfo
  {journal} {Phys. Rev. Lett.}\ }\textbf {\bibinfo {volume} {90}},\ \bibinfo
  {pages} {242001} (\bibinfo {year} {2003})},\ \Eprint
  {http://arxiv.org/abs/hep-ex/0304021} {arXiv:hep-ex/0304021 [hep-ex]}
  \BibitemShut {NoStop}%
\bibitem [{\citenamefont {Besson}\ \emph {et~al.}(2003)\citenamefont {Besson}
  \emph {et~al.}}]{Besson:2003cp}%
  \BibitemOpen
  \bibfield  {author} {\bibinfo {author} {\bibfnamefont {D.}~\bibnamefont
  {Besson}} \emph {et~al.} (\bibinfo {collaboration} {CLEO}),\ }\href {\doibase
  10.1103/PhysRevD.68.032002, 10.1103/PhysRevD.75.119908} {\bibfield  {journal}
  {\bibinfo  {journal} {Phys. Rev.}\ }\textbf {\bibinfo {volume} {D68}},\
  \bibinfo {pages} {032002} (\bibinfo {year} {2003})},\ \bibinfo {note}
  {[Erratum: Phys. Rev.D75,119908(2007)]},\ \Eprint
  {http://arxiv.org/abs/hep-ex/0305100} {arXiv:hep-ex/0305100 [hep-ex]}
  \BibitemShut {NoStop}%
\bibitem [{\citenamefont {Chen}\ \emph {et~al.}(2017)\citenamefont {Chen},
  \citenamefont {Chen}, \citenamefont {Liu}, \citenamefont {Liu},\ and\
  \citenamefont {Zhu}}]{Chen:2016spr}%
  \BibitemOpen
  \bibfield  {author} {\bibinfo {author} {\bibfnamefont {H.-X.}\ \bibnamefont
  {Chen}}, \bibinfo {author} {\bibfnamefont {W.}~\bibnamefont {Chen}}, \bibinfo
  {author} {\bibfnamefont {X.}~\bibnamefont {Liu}}, \bibinfo {author}
  {\bibfnamefont {Y.-R.}\ \bibnamefont {Liu}}, \ and\ \bibinfo {author}
  {\bibfnamefont {S.-L.}\ \bibnamefont {Zhu}},\ }\href {\doibase
  10.1088/1361-6633/aa6420} {\bibfield  {journal} {\bibinfo  {journal} {Rept.
  Prog. Phys.}\ }\textbf {\bibinfo {volume} {80}},\ \bibinfo {pages} {076201}
  (\bibinfo {year} {2017})},\ \Eprint {http://arxiv.org/abs/1609.08928}
  {arXiv:1609.08928 [hep-ph]} \BibitemShut {NoStop}%
\bibitem [{\citenamefont {van Beveren}\ and\ \citenamefont
  {Rupp}(2003)}]{vanBeveren:2003kd}%
  \BibitemOpen
  \bibfield  {author} {\bibinfo {author} {\bibfnamefont {E.}~\bibnamefont {van
  Beveren}}\ and\ \bibinfo {author} {\bibfnamefont {G.}~\bibnamefont {Rupp}},\
  }\href {\doibase 10.1103/PhysRevLett.91.012003} {\bibfield  {journal}
  {\bibinfo  {journal} {Phys. Rev. Lett.}\ }\textbf {\bibinfo {volume} {91}},\
  \bibinfo {pages} {012003} (\bibinfo {year} {2003})},\ \Eprint
  {http://arxiv.org/abs/hep-ph/0305035} {arXiv:hep-ph/0305035 [hep-ph]}
  \BibitemShut {NoStop}%
\bibitem [{\citenamefont {Dai}\ \emph {et~al.}(2003)\citenamefont {Dai},
  \citenamefont {Huang}, \citenamefont {Liu},\ and\ \citenamefont
  {Zhu}}]{Dai:2003yg}%
  \BibitemOpen
  \bibfield  {author} {\bibinfo {author} {\bibfnamefont {Y.-B.}\ \bibnamefont
  {Dai}}, \bibinfo {author} {\bibfnamefont {C.-S.}\ \bibnamefont {Huang}},
  \bibinfo {author} {\bibfnamefont {C.}~\bibnamefont {Liu}}, \ and\ \bibinfo
  {author} {\bibfnamefont {S.-L.}\ \bibnamefont {Zhu}},\ }\href {\doibase
  10.1103/PhysRevD.68.114011} {\bibfield  {journal} {\bibinfo  {journal} {Phys.
  Rev.}\ }\textbf {\bibinfo {volume} {D68}},\ \bibinfo {pages} {114011}
  (\bibinfo {year} {2003})},\ \Eprint {http://arxiv.org/abs/hep-ph/0306274}
  {arXiv:hep-ph/0306274 [hep-ph]} \BibitemShut {NoStop}%
\bibitem [{\citenamefont {Guo}\ \emph {et~al.}(2006)\citenamefont {Guo},
  \citenamefont {Shen}, \citenamefont {Chiang}, \citenamefont {Ping},\ and\
  \citenamefont {Zou}}]{Guo:2006fu}%
  \BibitemOpen
  \bibfield  {author} {\bibinfo {author} {\bibfnamefont {F.-K.}\ \bibnamefont
  {Guo}}, \bibinfo {author} {\bibfnamefont {P.-N.}\ \bibnamefont {Shen}},
  \bibinfo {author} {\bibfnamefont {H.-C.}\ \bibnamefont {Chiang}}, \bibinfo
  {author} {\bibfnamefont {R.-G.}\ \bibnamefont {Ping}}, \ and\ \bibinfo
  {author} {\bibfnamefont {B.-S.}\ \bibnamefont {Zou}},\ }\href {\doibase
  10.1016/j.physletb.2006.08.064} {\bibfield  {journal} {\bibinfo  {journal}
  {Phys. Lett.}\ }\textbf {\bibinfo {volume} {B641}},\ \bibinfo {pages} {278}
  (\bibinfo {year} {2006})},\ \Eprint {http://arxiv.org/abs/hep-ph/0603072}
  {arXiv:hep-ph/0603072 [hep-ph]} \BibitemShut {NoStop}%
\bibitem [{\citenamefont {Lang}\ \emph {et~al.}(2014)\citenamefont {Lang},
  \citenamefont {Leskovec}, \citenamefont {Mohler}, \citenamefont {Prelovsek},\
  and\ \citenamefont {Woloshyn}}]{Lang:2014yfa}%
  \BibitemOpen
  \bibfield  {author} {\bibinfo {author} {\bibfnamefont {C.~B.}\ \bibnamefont
  {Lang}}, \bibinfo {author} {\bibfnamefont {L.}~\bibnamefont {Leskovec}},
  \bibinfo {author} {\bibfnamefont {D.}~\bibnamefont {Mohler}}, \bibinfo
  {author} {\bibfnamefont {S.}~\bibnamefont {Prelovsek}}, \ and\ \bibinfo
  {author} {\bibfnamefont {R.~M.}\ \bibnamefont {Woloshyn}},\ }\href {\doibase
  10.1103/PhysRevD.90.034510} {\bibfield  {journal} {\bibinfo  {journal} {Phys.
  Rev.}\ }\textbf {\bibinfo {volume} {D90}},\ \bibinfo {pages} {034510}
  (\bibinfo {year} {2014})},\ \Eprint {http://arxiv.org/abs/1403.8103}
  {arXiv:1403.8103 [hep-lat]} \BibitemShut {NoStop}%
\bibitem [{\citenamefont {Liu}\ \emph {et~al.}(2009)\citenamefont {Liu},
  \citenamefont {Liu},\ and\ \citenamefont {Zhu}}]{Liu:2009uz}%
  \BibitemOpen
  \bibfield  {author} {\bibinfo {author} {\bibfnamefont {Y.-R.}\ \bibnamefont
  {Liu}}, \bibinfo {author} {\bibfnamefont {X.}~\bibnamefont {Liu}}, \ and\
  \bibinfo {author} {\bibfnamefont {S.-L.}\ \bibnamefont {Zhu}},\ }\href
  {\doibase 10.1103/PhysRevD.79.094026} {\bibfield  {journal} {\bibinfo
  {journal} {Phys. Rev.}\ }\textbf {\bibinfo {volume} {D79}},\ \bibinfo {pages}
  {094026} (\bibinfo {year} {2009})},\ \Eprint {http://arxiv.org/abs/0904.1770}
  {arXiv:0904.1770 [hep-ph]} \BibitemShut {NoStop}%
\bibitem [{\citenamefont {Liu}\ \emph {et~al.}(2011{\natexlab{a}})\citenamefont
  {Liu}, \citenamefont {Liu}, \citenamefont {Liu},\ and\ \citenamefont
  {Zhu}}]{Liu:2011mi}%
  \BibitemOpen
  \bibfield  {author} {\bibinfo {author} {\bibfnamefont {Z.-W.}\ \bibnamefont
  {Liu}}, \bibinfo {author} {\bibfnamefont {Y.-R.}\ \bibnamefont {Liu}},
  \bibinfo {author} {\bibfnamefont {X.}~\bibnamefont {Liu}}, \ and\ \bibinfo
  {author} {\bibfnamefont {S.-L.}\ \bibnamefont {Zhu}},\ }\href {\doibase
  10.1103/PhysRevD.84.034002} {\bibfield  {journal} {\bibinfo  {journal} {Phys.
  Rev.}\ }\textbf {\bibinfo {volume} {D84}},\ \bibinfo {pages} {034002}
  (\bibinfo {year} {2011}{\natexlab{a}})},\ \Eprint
  {http://arxiv.org/abs/1104.2726} {arXiv:1104.2726 [hep-ph]} \BibitemShut
  {NoStop}%
\bibitem [{\citenamefont {Guo}\ \emph {et~al.}(2009)\citenamefont {Guo},
  \citenamefont {Hanhart},\ and\ \citenamefont {Meissner}}]{Guo:2009ct}%
  \BibitemOpen
  \bibfield  {author} {\bibinfo {author} {\bibfnamefont {F.-K.}\ \bibnamefont
  {Guo}}, \bibinfo {author} {\bibfnamefont {C.}~\bibnamefont {Hanhart}}, \ and\
  \bibinfo {author} {\bibfnamefont {U.-G.}\ \bibnamefont {Meissner}},\ }\href
  {\doibase 10.1140/epja/i2009-10762-1} {\bibfield  {journal} {\bibinfo
  {journal} {Eur. Phys. J.}\ }\textbf {\bibinfo {volume} {A40}},\ \bibinfo
  {pages} {171} (\bibinfo {year} {2009})},\ \Eprint
  {http://arxiv.org/abs/0901.1597} {arXiv:0901.1597 [hep-ph]} \BibitemShut
  {NoStop}%
\bibitem [{\citenamefont {Geng}\ \emph {et~al.}(2010)\citenamefont {Geng},
  \citenamefont {Kaiser}, \citenamefont {Martin-Camalich},\ and\ \citenamefont
  {Weise}}]{Geng:2010vw}%
  \BibitemOpen
  \bibfield  {author} {\bibinfo {author} {\bibfnamefont {L.~S.}\ \bibnamefont
  {Geng}}, \bibinfo {author} {\bibfnamefont {N.}~\bibnamefont {Kaiser}},
  \bibinfo {author} {\bibfnamefont {J.}~\bibnamefont {Martin-Camalich}}, \ and\
  \bibinfo {author} {\bibfnamefont {W.}~\bibnamefont {Weise}},\ }\href
  {\doibase 10.1103/PhysRevD.82.054022} {\bibfield  {journal} {\bibinfo
  {journal} {Phys. Rev.}\ }\textbf {\bibinfo {volume} {D82}},\ \bibinfo {pages}
  {054022} (\bibinfo {year} {2010})},\ \Eprint {http://arxiv.org/abs/1008.0383}
  {arXiv:1008.0383 [hep-ph]} \BibitemShut {NoStop}%
\bibitem [{\citenamefont {Altenbuchinger}\ \emph {et~al.}(2014)\citenamefont
  {Altenbuchinger}, \citenamefont {Geng},\ and\ \citenamefont
  {Weise}}]{Altenbuchinger:2013vwa}%
  \BibitemOpen
  \bibfield  {author} {\bibinfo {author} {\bibfnamefont {M.}~\bibnamefont
  {Altenbuchinger}}, \bibinfo {author} {\bibfnamefont {L.~S.}\ \bibnamefont
  {Geng}}, \ and\ \bibinfo {author} {\bibfnamefont {W.}~\bibnamefont {Weise}},\
  }\href {\doibase 10.1103/PhysRevD.89.014026} {\bibfield  {journal} {\bibinfo
  {journal} {Phys. Rev.}\ }\textbf {\bibinfo {volume} {D89}},\ \bibinfo {pages}
  {014026} (\bibinfo {year} {2014})},\ \Eprint {http://arxiv.org/abs/1309.4743}
  {arXiv:1309.4743 [hep-ph]} \BibitemShut {NoStop}%
\bibitem [{\citenamefont {Liu}\ \emph {et~al.}(2013)\citenamefont {Liu},
  \citenamefont {Orginos}, \citenamefont {Guo}, \citenamefont {Hanhart},\ and\
  \citenamefont {Meissner}}]{Liu:2012zya}%
  \BibitemOpen
  \bibfield  {author} {\bibinfo {author} {\bibfnamefont {L.}~\bibnamefont
  {Liu}}, \bibinfo {author} {\bibfnamefont {K.}~\bibnamefont {Orginos}},
  \bibinfo {author} {\bibfnamefont {F.-K.}\ \bibnamefont {Guo}}, \bibinfo
  {author} {\bibfnamefont {C.}~\bibnamefont {Hanhart}}, \ and\ \bibinfo
  {author} {\bibfnamefont {U.-G.}\ \bibnamefont {Meissner}},\ }\href {\doibase
  10.1103/PhysRevD.87.014508} {\bibfield  {journal} {\bibinfo  {journal} {Phys.
  Rev.}\ }\textbf {\bibinfo {volume} {D87}},\ \bibinfo {pages} {014508}
  (\bibinfo {year} {2013})},\ \Eprint {http://arxiv.org/abs/1208.4535}
  {arXiv:1208.4535 [hep-lat]} \BibitemShut {NoStop}%
\bibitem [{\citenamefont {Mohler}\ \emph {et~al.}(2013)\citenamefont {Mohler},
  \citenamefont {Lang}, \citenamefont {Leskovec}, \citenamefont {Prelovsek},\
  and\ \citenamefont {Woloshyn}}]{Mohler:2013rwa}%
  \BibitemOpen
  \bibfield  {author} {\bibinfo {author} {\bibfnamefont {D.}~\bibnamefont
  {Mohler}}, \bibinfo {author} {\bibfnamefont {C.~B.}\ \bibnamefont {Lang}},
  \bibinfo {author} {\bibfnamefont {L.}~\bibnamefont {Leskovec}}, \bibinfo
  {author} {\bibfnamefont {S.}~\bibnamefont {Prelovsek}}, \ and\ \bibinfo
  {author} {\bibfnamefont {R.~M.}\ \bibnamefont {Woloshyn}},\ }\href {\doibase
  10.1103/PhysRevLett.111.222001} {\bibfield  {journal} {\bibinfo  {journal}
  {Phys. Rev. Lett.}\ }\textbf {\bibinfo {volume} {111}},\ \bibinfo {pages}
  {222001} (\bibinfo {year} {2013})},\ \Eprint {http://arxiv.org/abs/1308.3175}
  {arXiv:1308.3175 [hep-lat]} \BibitemShut {NoStop}%
\bibitem [{\citenamefont {Moir}\ \emph {et~al.}(2016)\citenamefont {Moir},
  \citenamefont {Peardon}, \citenamefont {Ryan}, \citenamefont {Thomas},\ and\
  \citenamefont {Wilson}}]{Moir:2016srx}%
  \BibitemOpen
  \bibfield  {author} {\bibinfo {author} {\bibfnamefont {G.}~\bibnamefont
  {Moir}}, \bibinfo {author} {\bibfnamefont {M.}~\bibnamefont {Peardon}},
  \bibinfo {author} {\bibfnamefont {S.~M.}\ \bibnamefont {Ryan}}, \bibinfo
  {author} {\bibfnamefont {C.~E.}\ \bibnamefont {Thomas}}, \ and\ \bibinfo
  {author} {\bibfnamefont {D.~J.}\ \bibnamefont {Wilson}},\ }\href {\doibase
  10.1007/JHEP10(2016)011} {\bibfield  {journal} {\bibinfo  {journal} {JHEP}\
  }\textbf {\bibinfo {volume} {10}},\ \bibinfo {pages} {011} (\bibinfo {year}
  {2016})},\ \Eprint {http://arxiv.org/abs/1607.07093} {arXiv:1607.07093
  [hep-lat]} \BibitemShut {NoStop}%
\bibitem [{\citenamefont {Guo}\ \emph {et~al.}(2019)\citenamefont {Guo},
  \citenamefont {Liu}, \citenamefont {Meißner}, \citenamefont {Oller},\ and\
  \citenamefont {Rusetsky}}]{Guo:2018tjx}%
  \BibitemOpen
  \bibfield  {author} {\bibinfo {author} {\bibfnamefont {Z.-H.}\ \bibnamefont
  {Guo}}, \bibinfo {author} {\bibfnamefont {L.}~\bibnamefont {Liu}}, \bibinfo
  {author} {\bibfnamefont {U.-G.}\ \bibnamefont {Meißner}}, \bibinfo {author}
  {\bibfnamefont {J.~A.}\ \bibnamefont {Oller}}, \ and\ \bibinfo {author}
  {\bibfnamefont {A.}~\bibnamefont {Rusetsky}},\ }\href {\doibase
  10.1140/epjc/s10052-018-6518-1} {\bibfield  {journal} {\bibinfo  {journal}
  {Eur. Phys. J.}\ }\textbf {\bibinfo {volume} {C79}},\ \bibinfo {pages} {13}
  (\bibinfo {year} {2019})},\ \Eprint {http://arxiv.org/abs/1811.05585}
  {arXiv:1811.05585 [hep-ph]} \BibitemShut {NoStop}%
\bibitem [{\citenamefont {Falk}(1992)}]{Falk:1991nq}%
  \BibitemOpen
  \bibfield  {author} {\bibinfo {author} {\bibfnamefont {A.~F.}\ \bibnamefont
  {Falk}},\ }\href {\doibase 10.1016/0550-3213(92)90004-U} {\bibfield
  {journal} {\bibinfo  {journal} {Nucl. Phys.}\ }\textbf {\bibinfo {volume}
  {B378}},\ \bibinfo {pages} {79} (\bibinfo {year} {1992})}\BibitemShut
  {NoStop}%
\bibitem [{\citenamefont {Sun}\ \emph {et~al.}(2015)\citenamefont {Sun},
  \citenamefont {Liu}, \citenamefont {Liu},\ and\ \citenamefont
  {Zhu}}]{Sun:2014aya}%
  \BibitemOpen
  \bibfield  {author} {\bibinfo {author} {\bibfnamefont {Z.-F.}\ \bibnamefont
  {Sun}}, \bibinfo {author} {\bibfnamefont {Z.-W.}\ \bibnamefont {Liu}},
  \bibinfo {author} {\bibfnamefont {X.}~\bibnamefont {Liu}}, \ and\ \bibinfo
  {author} {\bibfnamefont {S.-L.}\ \bibnamefont {Zhu}},\ }\href {\doibase
  10.1103/PhysRevD.91.094030} {\bibfield  {journal} {\bibinfo  {journal} {Phys.
  Rev.}\ }\textbf {\bibinfo {volume} {D91}},\ \bibinfo {pages} {094030}
  (\bibinfo {year} {2015})},\ \Eprint {http://arxiv.org/abs/1411.2117}
  {arXiv:1411.2117 [hep-ph]} \BibitemShut {NoStop}%
\bibitem [{\citenamefont {Fettes}\ and\ \citenamefont
  {Meissner}(2001)}]{Fettes:2000bb}%
  \BibitemOpen
  \bibfield  {author} {\bibinfo {author} {\bibfnamefont {N.}~\bibnamefont
  {Fettes}}\ and\ \bibinfo {author} {\bibfnamefont {U.~G.}\ \bibnamefont
  {Meissner}},\ }\href {\doibase 10.1016/S0375-9474(00)00368-7} {\bibfield
  {journal} {\bibinfo  {journal} {Nucl. Phys.}\ }\textbf {\bibinfo {volume}
  {A679}},\ \bibinfo {pages} {629} (\bibinfo {year} {2001})},\ \Eprint
  {http://arxiv.org/abs/hep-ph/0006299} {arXiv:hep-ph/0006299 [hep-ph]}
  \BibitemShut {NoStop}%
\bibitem [{\citenamefont {Liu}\ \emph {et~al.}(2011{\natexlab{b}})\citenamefont
  {Liu}, \citenamefont {Liu},\ and\ \citenamefont {Zhu}}]{Liu:2010bw}%
  \BibitemOpen
  \bibfield  {author} {\bibinfo {author} {\bibfnamefont {Z.-W.}\ \bibnamefont
  {Liu}}, \bibinfo {author} {\bibfnamefont {Y.-R.}\ \bibnamefont {Liu}}, \ and\
  \bibinfo {author} {\bibfnamefont {S.-L.}\ \bibnamefont {Zhu}},\ }\href
  {\doibase 10.1103/PhysRevD.83.034004} {\bibfield  {journal} {\bibinfo
  {journal} {Phys. Rev.}\ }\textbf {\bibinfo {volume} {D83}},\ \bibinfo {pages}
  {034004} (\bibinfo {year} {2011}{\natexlab{b}})},\ \Eprint
  {http://arxiv.org/abs/1011.3613} {arXiv:1011.3613 [hep-ph]} \BibitemShut
  {NoStop}%
\bibitem [{\citenamefont {Meng}\ \emph
  {et~al.}(2018{\natexlab{b}})\citenamefont {Meng}, \citenamefont {Wang},
  \citenamefont {Leng}, \citenamefont {Liu},\ and\ \citenamefont
  {Zhu}}]{Meng:2018gan}%
  \BibitemOpen
  \bibfield  {author} {\bibinfo {author} {\bibfnamefont {L.}~\bibnamefont
  {Meng}}, \bibinfo {author} {\bibfnamefont {G.-J.}\ \bibnamefont {Wang}},
  \bibinfo {author} {\bibfnamefont {C.-Z.}\ \bibnamefont {Leng}}, \bibinfo
  {author} {\bibfnamefont {Z.-W.}\ \bibnamefont {Liu}}, \ and\ \bibinfo
  {author} {\bibfnamefont {S.-L.}\ \bibnamefont {Zhu}},\ }\href {\doibase
  10.1103/PhysRevD.98.094013} {\bibfield  {journal} {\bibinfo  {journal} {Phys.
  Rev.}\ }\textbf {\bibinfo {volume} {D98}},\ \bibinfo {pages} {094013}
  (\bibinfo {year} {2018}{\natexlab{b}})},\ \Eprint
  {http://arxiv.org/abs/1805.09580} {arXiv:1805.09580 [hep-ph]} \BibitemShut
  {NoStop}%
\bibitem [{\citenamefont {Grinstein}\ \emph {et~al.}(1992)\citenamefont
  {Grinstein}, \citenamefont {Jenkins}, \citenamefont {Manohar}, \citenamefont
  {Savage},\ and\ \citenamefont {Wise}}]{Grinstein:1992qt}%
  \BibitemOpen
  \bibfield  {author} {\bibinfo {author} {\bibfnamefont {B.}~\bibnamefont
  {Grinstein}}, \bibinfo {author} {\bibfnamefont {E.~E.}\ \bibnamefont
  {Jenkins}}, \bibinfo {author} {\bibfnamefont {A.~V.}\ \bibnamefont
  {Manohar}}, \bibinfo {author} {\bibfnamefont {M.~J.}\ \bibnamefont {Savage}},
  \ and\ \bibinfo {author} {\bibfnamefont {M.~B.}\ \bibnamefont {Wise}},\
  }\href {\doibase 10.1016/0550-3213(92)90248-A} {\bibfield  {journal}
  {\bibinfo  {journal} {Nucl. Phys.}\ }\textbf {\bibinfo {volume} {B380}},\
  \bibinfo {pages} {369} (\bibinfo {year} {1992})},\ \Eprint
  {http://arxiv.org/abs/hep-ph/9204207} {arXiv:hep-ph/9204207 [hep-ph]}
  \BibitemShut {NoStop}%
\bibitem [{\citenamefont {Ebert}\ \emph {et~al.}(2002)\citenamefont {Ebert},
  \citenamefont {Faustov}, \citenamefont {Galkin},\ and\ \citenamefont
  {Martynenko}}]{Ebert:2002ig}%
  \BibitemOpen
  \bibfield  {author} {\bibinfo {author} {\bibfnamefont {D.}~\bibnamefont
  {Ebert}}, \bibinfo {author} {\bibfnamefont {R.~N.}\ \bibnamefont {Faustov}},
  \bibinfo {author} {\bibfnamefont {V.~O.}\ \bibnamefont {Galkin}}, \ and\
  \bibinfo {author} {\bibfnamefont {A.~P.}\ \bibnamefont {Martynenko}},\ }\href
  {\doibase 10.1103/PhysRevD.66.014008} {\bibfield  {journal} {\bibinfo
  {journal} {Phys. Rev.}\ }\textbf {\bibinfo {volume} {D66}},\ \bibinfo {pages}
  {014008} (\bibinfo {year} {2002})},\ \Eprint
  {http://arxiv.org/abs/hep-ph/0201217} {arXiv:hep-ph/0201217 [hep-ph]}
  \BibitemShut {NoStop}%
\bibitem [{\citenamefont {Tanabashi}\ \emph {et~al.}(2018)\citenamefont
  {Tanabashi} \emph {et~al.}}]{Tanabashi:2018oca}%
  \BibitemOpen
  \bibfield  {author} {\bibinfo {author} {\bibfnamefont {M.}~\bibnamefont
  {Tanabashi}} \emph {et~al.} (\bibinfo {collaboration} {Particle Data
  Group}),\ }\href {\doibase 10.1103/PhysRevD.98.030001} {\bibfield  {journal}
  {\bibinfo  {journal} {Phys. Rev.}\ }\textbf {\bibinfo {volume} {D98}},\
  \bibinfo {pages} {030001} (\bibinfo {year} {2018})}\BibitemShut {NoStop}%
\bibitem [{\citenamefont {Ahmed}\ \emph {et~al.}(2001)\citenamefont {Ahmed}
  \emph {et~al.}}]{Ahmed:2001xc}%
  \BibitemOpen
  \bibfield  {author} {\bibinfo {author} {\bibfnamefont {S.}~\bibnamefont
  {Ahmed}} \emph {et~al.} (\bibinfo {collaboration} {CLEO}),\ }\href {\doibase
  10.1103/PhysRevLett.87.251801} {\bibfield  {journal} {\bibinfo  {journal}
  {Phys. Rev. Lett.}\ }\textbf {\bibinfo {volume} {87}},\ \bibinfo {pages}
  {251801} (\bibinfo {year} {2001})},\ \Eprint
  {http://arxiv.org/abs/hep-ex/0108013} {arXiv:hep-ex/0108013 [hep-ex]}
  \BibitemShut {NoStop}%
\bibitem [{\citenamefont {Liu}\ \emph {et~al.}(2008)\citenamefont {Liu},
  \citenamefont {Lin},\ and\ \citenamefont {Orginos}}]{Liu:2008rza}%
  \BibitemOpen
  \bibfield  {author} {\bibinfo {author} {\bibfnamefont {L.}~\bibnamefont
  {Liu}}, \bibinfo {author} {\bibfnamefont {H.-W.}\ \bibnamefont {Lin}}, \ and\
  \bibinfo {author} {\bibfnamefont {K.}~\bibnamefont {Orginos}},\ }\bibfield
  {booktitle} {\emph {\bibinfo {booktitle} {{Proceedings, 26th International
  Symposium on Lattice field theory (Lattice 2008): Williamsburg, USA, July
  14-19, 2008}}},\ }\href {\doibase 10.22323/1.066.0112} {\bibfield  {journal}
  {\bibinfo  {journal} {PoS}\ }\textbf {\bibinfo {volume} {LATTICE2008}},\
  \bibinfo {pages} {112} (\bibinfo {year} {2008})},\ \Eprint
  {http://arxiv.org/abs/0810.5412} {arXiv:0810.5412 [hep-lat]} \BibitemShut
  {NoStop}%
\end{thebibliography}%

\end{document}